\documentclass[preprint,10pt,APS]{revtex4}%
\usepackage{amsmath}%
\usepackage{amsfonts}%
\usepackage{amssymb}%
\usepackage[dvips]{graphicx}
\usepackage{color}
\usepackage{psfrag}
\usepackage{epic,eepic}

\def\kcor#1{#1}
\begin{document}

\title{Grand Partition Functions of Little Matrix Models with ABCD}

\author{Hironobu Kihara}
\affiliation{Korea Institute for Advanced Study\\
207-43 Cheongnyangni 2-dong, Dongdaemun-gu, Seoul 130-722, Republic of Korea}

\date{March 28, 2008}
\preprint{KIAS-P08027}

\begin{abstract}
Itoyama-Tokura type USp matrix model is discussed. 
Non-Abelian Berry's phases in a T-dualized model of IT model were reconsidered. 
These phases describe the higher dimensional monopoles; Yang monopole and nine-dimensional monopole. 
They are described by the connections of the BPST instanton on $S^4$ and the Tchrakian-GKS instanton on $S^8$, respectively.   

As a preparation to understand their effect in original zero-dimensional model,
we consider partition function of simplified matrix models. 
We compute partition functions of SU, SO and USp reduced matrix models. 
Groups SO and USp appear in low energy effective theories of string against orientifold background. 
In this evaluation we chose different poles from that of Moore-Nekrasov-Shatashvili and our previous result. 

The position of poles explain branes' and the orientifold's configurations. 
There is a brane which is sitting on the orientifold in the SO($2N$) model, while in USp($2N$) and SO$(2N+1)$ model there are no branes on the orientifold. 
The grand partition functions of these models are considered. 
They follow to linear second order ordinary differential equations and their singularities are $q=0,\infty$. 
Their solutions can be analytically continued to whole $q$ plane.   
We show the expectation values of the number $N$ of A and C cases as examples. 
 There is an ambiguity coming from the problem on sign. 
Grand partition functions with minus sign give effective actions which have cusp singularities. 
\end{abstract}

\maketitle

\section{Introduction}
In 1997, Itoyama and Tokura considered a matrix model which we call USp matrix model. The model was studied in order to understand the dynamics of string against the orientifold background \cite{Itoyama:1998et}. There is another USp matrix model by S.J. Rey and N.G. Kim \cite{Kim:1997gh} and their work must be important for us.  
Because the eleven-dimensional minimal coupling supergravity theory contains rank-three antisymmetric tensor fields, the theory has been considered as the low-energy effective theory of the extended object ``{\it membrane}" whose trajectories of movement are, in general, $2+1$-dimensional space.  
The mass spectrum of the supersymmetric membrane is described by the Hamiltonian of the dimensionally reduced model of the ten-dimensional supersymmetric Yang-Mills theory to one dimension. 

Type IIA superstring theory has infinitely many BPS states which couple with the Ramond-Ramond one-form and these BPS states are identified with 
configurations of D-particles. 
The tower consisting of these configurations of D-particles can be interpreted into the Kaluza-Klein modes accompanied by the circular compactification along the eleventh direction. 
This viewpoint leads people to the discussion on the duality between the theory of membrane and IIA superstring theory \cite{Duff:1987bx,Banks:1996vh}. 
The matrix theory has the time direction and it is not clear whether the model has covariance or not. 
Ishibashi, Kawai, Kitazawa, and Tsuchiya have proposed a matrix model whose massless spectrum is the same as that of type IIB string theory \cite{Ishibashi:1996xs}. 
It was conjectured that the large-$N$ reduced model of ten-dimensional super Yang-Mills theory can be regarded as a constructive definition of string theory. 
\begin{align*}
Z &= \sum_{N=0}^{\infty} \int [dX^{(N)} d \Psi^{(N)} ] \exp(-S_N) ~,&
S_N &= \alpha \left( - \frac{1}{4} {\rm Tr} [X^{(N)}_{\hat{\mu}} , X^{(N)}_{\hat{\nu}} ]^2 - \frac{1}{2} {\rm Tr} \bar{\Psi}^{(N)} \Gamma^{\hat{\mu}} [ X^{(N)}_{\hat{\mu}} , \Psi^{(N)} ]     \right)   + \beta {\rm Tr} {\bf 1}~,
\end{align*}
where $X^{(N)}_{\hat{\mu}}$ and $\Psi^{(N)}$ are bosonic and fermionic $ N \times N$ Hermitian matrices, respectively and $\hat{\mu}=0,1 , \cdots, 9$. $\alpha = 1/g^2$ is the inverse of string tension or gauge coupling and $\beta$ is the chemical potential which is needed for insertion of one instanton. 

 If in fact the IKKT model unifies string theories, the model should not depend on the background.   
The background independence is the most significant implication in the inclusion of quantum gravity in string theory. The geometry of space-time should not be set up a priori, rather it is generated by a highly nonperturbative effect, the condensation of strings.
 Therefore understanding of the background independence is promisingly the key ingredient to seek the underlying principle of nonperturbative string theory. 

We can include another type of background. 
Orientifolds are generalized orbifolds. In the orbifold construction, discrete internal symmetries of the world-sheet theory are gauged. 
In the orientifold, products of internal symmetries with world-sheet parity reversal are also gauged. 
Roughly speaking, these symmetries yield restrictions of gauge groups to unitary symplectic (USp) or orthogonal (SO) groups. 

\kcor{In this article, we will consider partition functions of reduced matrix models. 
Several people evaluate the quantity. The quantity relates to the Witten index \cite{Witten:1982df} of matrix quantum mechanics and to the dynamics of D-particles in the context of M-theory \cite{Witten:1995im,Yi:1997eg,Moore:1998et,Nekrasov:2004vw}. 
Monte-Carlo simulation has been considered \cite{Krauth:1998xh,Ambjorn:2000dx,Oda:2000im} and the convergence of the integral was studied in
\cite{Austing:2001bd}. Moore-Nekrasov-Shatashvili obtained the result $1/N^2$ by using the deformation method \cite{Moore:1998et} and the Monte-Carlo result by 
Krauth-Staudacher \cite{Krauth:1998xh}  agree with the MNS result.} 
\kcor{We start from the MNS's contour integral and obtain the different result from the result $1/N^2$ for SU(N) case.  
There are four types of the partition functions $Z_{{\rm RMM}}, Z_{{\rm MC}}, Z_{{\rm MNS}}, Z_{{\rm X}}$.  
The true partition function of the reduced matrix model with suitable normalization factor is denoted $Z_{{\rm RMM}}$, the result of the Monte-Carlo simulation which might be used different normalization factor is $Z_{{\rm MC}}$, the MNS result is $Z_{{\rm MNS}}$ and ours is $Z_{{\rm X}}$. 
The partition function $Z_{{\rm MC}}$ is a finite list, however it agrees with $Z_{{\rm MNS}}$ for SU(2) and SU(3) cases. 
Let us start from doubt of the equality between $Z_{{\rm RMM}}$ and $Z_{{\rm MNS}}$. 
It implies that we do not believe the equivalence between $Z_{{\rm RMM}}$ and $Z_{{\rm MC}}$. 
For SU(2) case, our result agrees with their result $1/4$, while SU(3) case there exist the difference of factor 2. 
Two quantities $Z_{{\rm MNS}}$, $Z_{{\rm X}}$ are different with each other. 
The difference is occurred from the difference of the choice of the contour. 
In this paper, we will not discuss the justice of these results, but will discuss the meaning of the multiplicity.
Our resultant effective potential for SU series has slightly similar shape with a graph in \cite{Nishimura:2008ta}, 
however our independent variable is the expectation value of the matrix size and their variable is temperature.}

There are two kinds of bosonic sector of Type IIB superstring theory; i) {\bf NS-NS} sector: dilaton $\phi$, graviton $g_{\hat{\mu}\hat{\nu}}$, two-form field $B_{\hat{\mu}\hat{\nu}}^{\rm F}$; ii) {\bf R-R} sector: axion $\chi$, two-form field $B_{\hat{\mu}\hat{\nu}}^{\rm D}$, self-dual four-form field 
$A_{\hat{\mu}\hat{\nu}\hat{\rho}\hat{\tau}}^+$. 
The combination of these scalar fields {$\tau= \chi + i \exp(-\phi)$} is regarded as a modulus of torus fiber $T^2$. Then the doublet {$(B_{\hat{\mu}\hat{\nu}}^{\rm F}, B_{\hat{\mu}\hat{\nu}}^{\rm D})$} becomes a $SL(2,{\mathbb Z})$ multiplet. 
Therefore IIB superstring on a compactified space $B$ is considered as a compactification of F theory over $M$, where
 fiber bundle $ \pi : M \rightarrow B$ ($\pi^{-1}(z) \simeq T^2$) is defined by the moduli function $\tau(z)$. 
One famous example of this configuration is $M =K3$ and $B={\mathbb C}{\bf P}^1 \simeq S^2$. 
In this case, F-theory over $K3$ is dual to IIB over $S^2$. Here K3 is smooth Calabi-Yau  2-fold. 
Especially F theory over $K3$ near the orbifold limit of $K3$ describes the dynamics of orientifold of Type IIB over $T^2$. 
On the other hand, orientifold of Type IIB on $T^2$ is T-dual to Type I  on $T^2$. 
One Realization of $K3$ (not generic) is a quartic surface in ${\mathbb C}{\bf P}^3$; $V(f)=\{ Z \in {\mathbb C}{\bf P}^3 | f(Z)=0 , {\rm deg}f =4, f: \mbox{ homogeneous } \}$.
 The cohomology class is obtained by the K\"ahler 2-form on ${\mathbb C}{\bf P}^3$; 
$(1+J)^4/(1+4J) \sim 1 + 6 J^2$. From this, we can read off that the Euler number is equal to 24.

Let us talk about Itoyama-Tokura Model which is a kind of reduced matrix model. 
The model is written in terms of dimensional reduction of the $d=4,{\cal N}=2$ USp($2N$) supersymmetric gauge theory  to zero dimension. The model has i) one adjoint vectormultiplet;
ii) one anti-symmetric hypermultiplet;
iii) $N_f$ fundamental hypermultiplets.
The total action of this model consists of two parts; $S_{Tot}= S_{C} +  S_{O}$. Here the closed sector $S_{C}$ is obtained by orientifold projection from IKKT action and the open sector
$S_{O}$ is from space-time filling D-brane. 
Three representations of USp group appear. 
The adjoint and the antisymmetric representations are, in a sense, parts of Hermitian matrices;
\begin{align*}
u(2N) &= \{ i X \in M(2N , {\mathbb C} ) | X^{\dag} = X  \} ~,&
usp(2N) &= \{ i X \in u(2N) | X^T J + J X=0 \} \\
asym(2N) &= \{ i X \in u(2N) | X^T J - J X=0 \} ~,&
J&= \begin{pmatrix} 0 & I_N \\ - I_N & 0 \end{pmatrix}
\end{align*}
Therefore  $u(2N)$ is a direct sum of adjoint and antisymmetric;  $u(2N) \simeq usp(2N) \oplus asym(2N)$. 
Let $i X$ be an element of $usp(2N)$ and  $i Y$  be an element of $asym(2N)$. The block notation show us their construction
\begin{align*}
X &= \begin{pmatrix} A & B \\ B^{\dag} & - A^{T}\end{pmatrix}~,&
Y &= \begin{pmatrix} C & D \\ D^{\dag} &  C^{T}\end{pmatrix}~,
\end{align*}
where $A^{\dag} = A,~~ C^{\dag} = C,~~ B^T =B,~~ D^T = - D.$

The orientifold projection is described as follows.  
The ${\cal N}=4$ SU($2N$) supersymmetric Yang-Mills theory consists of four multiplets $V, \Phi_1, \Phi_2, \Phi_3$ which are all adjoint  representation of  SU($2N$). 
Let us project them into ${\cal N}=2$ USp($2N$) supersymmetric gauge theory with adjoint representations $V,\Phi_1$ and antisymmetric representations $\Phi_2, \Phi_3$.

We also add $N_f$ fundamental representations $Q_i, \tilde{Q}_i$ by hand. 
In other words, a ten-dimensional vector splits into six-dimensional vector and four scalars; 
$\hat{X}_{\hat{\mu}} \in su(2N)$ $\Rightarrow$ 
$X_{\mu} \in usp(2N)$ $(\mu=0,1,\cdots, 4,7) \oplus X_a \in asym(2N) $ $(a=5,6,8,9)$. 
Splitting of a Majorana-Weyl fermion 
$\hat{\Psi}$ which belongs to the SU($2N$) adjoint representation can be considered; $\Psi = \psi + \lambda$, where $\psi$ is a USp($2N$) adjoint eight component spinor and $\lambda$ is an  anti-symmetric eight component spinor. 
The action of the closed sector, $S_C$, is
\begin{align*}
S_C &= - \frac{1}{4g^2} {\rm Tr} [ X_{\hat{\mu}} , X_{\hat{\nu}}]^2  - \frac{1}{2g^2} {\rm Tr} \bar{\Psi} \Gamma^{\hat{\mu}} [ X_{\hat{\mu}} , \Psi ]. 
\end{align*}
Let us omit the explanation of the open sector $S_O$. In this model, the classical configuration is given by diagonal matrices $X_M$ and all fermions are set to zero; $\Psi=0$,~$Q,\tilde{Q}=0$. 
\begin{align*}
X_{\mu} &= {\rm diag} ( x_{\mu}^1 , \cdots , x_{\mu}^N , -x_{\mu}^1 , \cdots ,  - x_{\mu}^N)~,&
X_{a} &= {\rm diag} ( x_{a}^1 , \cdots , x_{a}^N , x_{a}^1 , \cdots , x_{a}^N)
\end{align*}
They form $N$ pairs of ten-dimensional vectors; $\{(x_{\mu}^i,x_a^i), (-x_{\mu}^i,x_a^i)\}_{i=1,\cdots, N}$. 
In each pair, one vector is a mirror image of the other with respect to the orientifold. 
In order to reveal the physics of the fermion of the fundamental matter, 
we computed the non-Abelian Berry's phases against the classical background \cite{Itoyama:1998uz,Chen:1998qb}. 
It was computed as the fermionic integration of the matrix model. In the paper, we did not consider the effect from the fermionic diagonal terms. In order to make the discussion precise, we should take account into the bilinear terms with respect to the fermionic diagonal elements, which are yielded from the quadratic completion of fermionic off-diagonal terms. 
In this paper we will not treat this problem. In fact such a term has influence on the integration of the bosonic off-diagonal degrees. 
Suppose that $X_{\hat{\mu}}$ is diagonal. We split the fermion into diagonal and off-diagonal terms; $\Psi = \Psi_D + \Psi_O$. As we mentioned above, we neglect $\Psi_D$. 
Fermionic part of the whole action is
\begin{align*}
S_F &= \frac{1}{2g^2} {\rm Tr} \bar{\Psi}_O \Gamma^{\hat{\mu}} ({\rm ad}X_{\hat{\mu}}) \Psi_O 
+ \sum_{i=1}^{n_f} \left\{ \bar{\chi}_i^{(I)} J( \gamma^{\mu} X_{\mu} + {{\cal M}_i} )  \chi_i^{(I)} + \bar{\chi}_i^{(II)} 
J( \gamma^{\mu} X_{\mu} + {{\cal M}_i} ) \chi_i^{(II)} \right\},
\end{align*}
where $\gamma^{\mu}$ are six-dimensional  Dirac matrices 
and the eight-component spinors $\chi_i^{(I)}, \chi_i^{(II)}$ are components of the fundamental hypermultiplets $Q_i, \tilde{Q}_i$, respectively. 
The $n_f \times n_f$ matrix ${\cal M}$ is the mass matrix of the fundamental matter and please permit to restrict our discussion in the case that the mass matrix ${\cal M}$ is diagonal. 
The off-diagonal part can be written in the components using roots and weights;
\begin{align*}
\Psi_O = \sum_{\alpha \in \Delta} \psi_{\alpha} T_{\alpha} + \sum_{\omega\in A'} \lambda_{\omega } T_{\omega}, 
\end{align*}
where $\Delta$ is the root system of $usp(2N)$  and 
$A'$ is a set of non-zero weights with respect to the anti-symmetric representation $asym(2N)$. 
$T_{\alpha}$ and $T_{\omega}$ are  generators of $usp(2N)$ and weight vector of $asym(2N)$, respectively. They are represented as unitary matrices.  
Let $H= {\rm diag}(h_1 , \cdots, h_N , - h_1 , \cdots, -h_N)$ be an element of a Cartan subalgebra. Then the dual basis is defined as ${\bf e}_i(H)=h_i$ $( i=1,2,\cdots ,N)$. 
\begin{align*}
\Delta &= \{ \pm ({\bf e}_i - {\bf e}_j ) , \pm ({\bf e}_i + {\bf e}_j ) , i < j , 2 {\bf e}_i, i=1,\cdots, N\}\\
A' &= \{ \pm ({\bf e}_i - {\bf e}_j ) , \pm ({\bf e}_i + {\bf e}_j ) , i < j \}
\end{align*}
Components $\psi_{\alpha}$, $\lambda_{\omega}$ from roots and weights which are explained like $\pm ({\bf e}_i - {\bf e}_j )$, $\pm ({\bf e}_i + {\bf e}_j )$ form sixteen-component spinors and components from $\pm 2{\bf e}_i$ form eight component spinors.  
In other words, the composition $( \psi_{\pm({\bf e}_i - {\bf e}_j )} , \lambda_{\pm({\bf e}_i - {\bf e}_j )})$ and $( \psi_{\pm({\bf e}_i + {\bf e}_j )} , \lambda_{\pm({\bf e}_i + {\bf e}_j )})$ form sixteen-component spinors, while $(\psi_{\pm 2 {\bf e}_i} , 0)$ are still eight-component spinors. 
Therefore we obtain two types of action;
i) $S_{I}[\xi , z] = \bar{\xi} \gamma^{\mu} z_{\mu} \xi$ where $\xi$ is a  8 component spinor,
 and $z$ is a  6-dimensional vector; ii) $S_{II}[\Xi , y] = \bar{\Xi} \Gamma^{\hat{\mu}} y_{\hat{\mu}} \Xi$ where $\Xi$ is a 16 component spinor and  $y$ is a 10-dimensional vector. 
However this model is a model on zero-dimensional space (a set of discrete points). 
Let us consider T-duality in order to ``make" time direction and let us concentrate on the ``one-particle state". Then Hamiltonian becomes 
\begin{align*}
H_{I}[z] &= \gamma^{\mu} z_{\mu}~,&H_{II}[y] &= \Gamma^{\hat{\mu}} y_{\hat{\mu}} ~.
\end{align*}
We set $y_0=0$ and $z_0=0$ and evaluate their Berry's phases in consideration of degeneracy. 
These Hamiltonians are higher dimensional generalizations of the system where spin couple to background magnetic field. In the case of $H_I[z]$, the background magnetic field are given by the Hodge dual of four form field strength and the corresponding gauge fields are three form field. 
Because of the degeneracy of these Hamiltonians, we obtain non-Abelian connections which are related to the generalized Monopoles. They are generalization of the Dirac monopole in five-dimensional space and nine-dimensional space. 
Berry's connections are 
\begin{align*}
\boldsymbol{z}&=(z_1,z_2,z_3,z_4,z_5)^t~,&\boldsymbol{y}&=(y_1,\cdots, z_9)^t~,&
R_z^2 &= \boldsymbol{z}^t \cdot \boldsymbol{z}~,& R_y^2 &= \boldsymbol{y}^t \cdot \boldsymbol{y} ~, 
\end{align*}
\begin{align}
Z &= \frac{1}{\tau} \left\{  z_4 + i (z_1 \sigma_1 + z_2 \sigma_2 + z_3 \sigma_3)  \right\}~,& \tau &= \sqrt{R_z^2 - z_5^2}~,&
A_{\rm Yang} &= \frac{\tau^2}{\tau^2 + \lambda^2} dZ Z^{-1}~,&\lambda &= R_z + z_5~.
\end{align}
Hypersurfaces defined by the condition $\lambda=const$ are hyperboloid and the hypersurfaces shrink to the Dirac string after taking limit $\lambda \rightarrow 0$. 
This five-dimensional monopole is called Yang monopole \cite{Yang:1977qv}.
If we transform this space to a space where the hypersurfaces $\lambda=const.$ become hyperplanes. 
There appears a mirage four plane where $\lambda=0$. 
The metric on the hyperplanes, which is induced from Euclidean metric on the original five-dimensional space, has non-trivial curvature. In fact, such a hyperboloid is a curved space.     
We omit the exhibition of the Tchrakian-GKS type solution. 
Thus the Hamiltonian $H_I$ gives BPST instanton connection on $S^4$ and the connection satisfies the self-dual equation; $F = \pm * F$. 
While remaining Hamiltonian $H_{II}$ gives the Tchrakian or GKS connection on $S^8$ and the connection satisfies the generalized self-dual equation $F \wedge F = \pm * F \wedge F$ \cite{Tchrakian:1978sf,Grossman:1984pi}.  
They are generalizations of Dirac monopole and the latter is generalization of the self duality.
In nine dimensional space, the shape of singularities look like a point. 
These branes which are denoted by solid lines come from fundamental multiplets and their distances are determined by the mass matrix. 

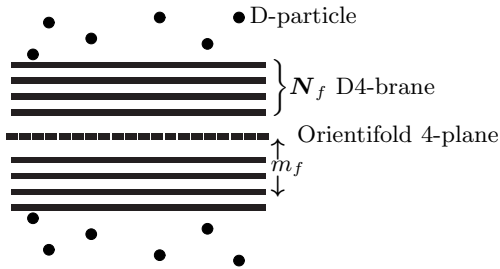
\begin{figure}[htb]
\begin{center}
\begin{picture}(180,100)(0,0)
\linethickness{2pt} 
\put(2,76){\line(96,0){96}}
\put(2,70){\line(96,0){96}}
\put(2,64){\line(96,0){96}}
\put(2,58){\line(96,0){96}}
\multiput(0,49)(5,0){20}{\line(4,0){4}}
\put(2,40){\line(96,0){96}}
\put(2,34){\line(96,0){96}}
\put(2,28){\line(96,0){96}}
\put(2,22){\line(96,0){96}}
\put(10,80){\circle*{4}}
\put(16,92){\circle*{4}}
\put(32,86){\circle*{4}}
\put(58,94){\circle*{4}}
\put(76,84){\circle*{4}}
\put(88,94){\circle*{4}}
\put(10,18){\circle*{4}}
\put(16,6){\circle*{4}}
\put(32,12){\circle*{4}}
\put(58,4){\circle*{4}}
\put(76,14){\circle*{4}}
\put(88,2){\circle*{4}}
\put(100,65){$\displaystyle \bigg\}  {\boldsymbol N_f}$ D4-brane}
\put(110,47){Orientifold 4-plane}
\put(92,92){D-particle}
\put(100,42){$\displaystyle {\boldsymbol \uparrow}$}
\put(100,36){$m_f$}
\put(100,28){$\displaystyle {\boldsymbol \downarrow}$}
\end{picture}
\caption{D-particles are from the Hamiltonian $H_{II}$ and orientifold and D4-branes are from $H_I$. }
\end{center}
\end{figure}

Our notation in \cite{Chen:1998qb} owed to \cite{Demler:1998pm} which was introduced by Itoyama. 
The non-Abelian generalization of Berry and Simon \cite{Berry:1984jv,Simon:1983mh} is argued by Wilczek and Zee \cite{Wilczek:1984dh}. 
The derivation of BPST instanton as Berry's phase is discussed in \cite{Levay:1991qw}. 

\section{Tchrakian's monopole}
Because the Yang monopole is a singular object like Dirac monopole, we cannot determine its mass or energy.  
In this section we will consider the Tchrakian's monopole \cite{Tchrakian:1978sf,Kihara:2004yz} as a possibility of the regularization of the Yang monopole. 
He constructed five-dimensional finite energy monopole solution which is an analogue of the 't Hooft-Polyakov monopole. 
The computation of the Berry's connection was done with adiabatic approximation. 
If the parameter $z$ is near the  origin, the adiabatic approximation is not valid because the gap of the spectrum of $H_I$ is not large enough. Such a gap is needed to avoid the transition between different energy eigen states. Therefore in order to clarify the behavior around the singular point, we need such a regularization. 

Let us work on the five-dimensional Euclidean Clifford algebra: $\{\gamma_a , \gamma_b \}= 2\delta_{ab}$,  where $a,b = 1,2,3,4,5$ and  $\sigma_{ab} = [\gamma_a , \gamma_b]/2$ are SO(5) generators. 
These gauge fields are brought together in one differential form which takes value in the Lie algebra; $A= (1/2) A_a^{bc} \sigma_{bc}dx^a$. 
The scalar fields are also represented as a matrix $\phi= \phi^a \gamma_a$.  
Field strength two form is $F=dA + g A^2$ where $g$ is a gauge coupling constant. The energy is defined as
\begin{align*}
E &= \int  {\rm Tr}\left\{ \frac{1}{8} (F \wedge F) \wedge *(F \wedge F) + \frac{1}{8} D\phi \wedge * D\phi + {\lambda} V(\phi) d^5x  \right\}~.
\end{align*}
The symbol $*$ is the Hodge dual operator with respect to the Euclidean metric on ${\mathbb R}^5$. 
From this energy, the Bogomol'nyi equation with the Prasad-Sommerfeld limit becomes $D\phi = \pm *(F \wedge F)$.
Let us consider the Hedge-Hog solution;
\begin{align*}
e&= x^a \gamma_a~,& r^2&= x^ax^a,&
A &= \frac{1-K(r)}{2g} ede,&
\phi&= H_0 U(r) e,
\end{align*}
where $H_0$ is the vacuum expectation value. Unknown functions are $K(r)$ and $U(r)$ and their boundary conditions are $U(0)=0,K(0)=1,U(\infty)=\pm 1, K(\infty)=0$. 
After change of the variable $a^3=2g^2 H_0 /3$, $s= {\rm ln}(ar)$,~$X(s)=K^2$, the equation becomes an autonomous differential equation:
\begin{align*}
Y(X) &= \frac{1}{X(1-X)} \frac{dX}{ds} ~,& X(1-X) Y \frac{dY}{dX} &= 2XY^2 + 3Y -2~.
\end{align*}
This equation is the Abel differential equation of the second kind. 
Let us put $Z=dY/dX$. Points $(X,Y,Z)$ are sitting on a surface defined by a quartic
 polynomial. The surface is singular and has a line singularity. 
The differential equation does not have the same property as that of Kovalevskaja. 
Fortunately numerical evaluation shows the existence of a flow which connect two boundary points. 
The effective theory against this background are written in terms of three form gauge field.  

\section{Partition Function of little SO and USp Matrix Model}
In this part we will compute the partition function of matrix models whose actions are given by the dimensional reduction of $d=4$ ${\cal N}=1$ supersymmetric Yang-Mills theories. Here gauge groups are classical groups and we list some properties of Lie groups in the appendix. 
Low dimensional counterparts of IKKT matrix model  have been discussed in 
\cite{Tomino:2003hb} and more. We use the terminology ``litte'' matrix model borrowed from them.  
The partition function of matrix models are considered in the context of the discussion about the existence of D-particle bound states;  \cite{Witten:1995im,Yi:1997eg,Moore:1998et,Nekrasov:2004vw}. 
They evaluated the Witten index of matrix quantum mechanics \cite{Witten:1982df}. 
We will follow to the equivariant deformation method which was used in \cite{Moore:1998et}. 
Once we performed the computation \cite{Itoyama:2006wb}. 
In this paper we will obtain different result from them. 

The partition function of a reduced matrix model is given as a matrix integral. Moore-Nekrasov-Shatashvili obtained the following result for $d=10$ ${\cal N}=1$ supersymmetric Yang-Mills theory.   
\begin{align*}
{\cal Z} &= \frac{1}{{\rm vol}(G)} \int [dX d \Psi ] \exp(-S)~,&
{\cal Z}_N &= \sum_{k | N} \frac{1}{k^2}~~.
\end{align*}
Here the summation with respect to $k$ is taken over the set which consists of all positive factors of $N$. 
In order to obtain these result, Moore-Nekrasov-Shatashvili used equivariant deformation method. 

Recently we have applied the deformation method to ${\cal N}=1, ~~d=4$ in the case of SO and USp gauge theories. In this paper, we retry the calculation and will study their grand partition functions. This work is a preparation to evaluate the integral ob the USp matrix model. 

Let us consider a model whose action is
\begin{align*}
{\cal S}[A_{N-1}] &= {\rm Tr} \left\{ - \frac{1}{4g^2} [v_m , v_n]^2 - \frac{1}{2g^2} \bar{\Lambda} \gamma^m [ v_m , \Lambda ]  \right\}~.
\end{align*}
Here the gauge group is SU($N$)$/{\mathbb Z}_N$ and suppose that the metric is Lorentzian $m,n=0,1,2,3$. 
This action is a low dimensional counterpart of the IKKT action. The signature of metric should cause a problem on its convergence. Suppose that the integration path of $\bar{\phi}$ is rotated. 
We will follow to the MNS's method and the matrix integral reduce to the residue calculus. 
 We will skip the detail of the derivation. 
The degrees of freedom are $v_m, \Lambda$ which are $N \times N$ Hermitian bosonic and fermionic matrices, respectively. 
Let $\{T_a\}_{a=1,\cdots, N^2-1}$ be generators of SU($N$). The matrices are expanded by these generators: $v_m = \sum v_m^a T_a$, $\Lambda = \sum \Lambda^a T_a$.  
The components $\Lambda^a$ are Majorana spinors: $\Lambda^a = (\lambda^a , \bar{\lambda}^a)^t$. 
 These are not fields, but constant matrices.  
Let us rearrange these matrices;
$
\phi = -(v_0 + v_3) ~, \bar{\phi} = v_0 - v_3~,
\lambda^1 = (\psi_1 + i \psi_2) /{\sqrt{2}}~,
\lambda^2 =  \eta /{2 \sqrt{2}} + i \chi$.
Let us use an auxiliary field $B$ and modify the action ${\cal S} \rightarrow {\cal S}[B]$. 
\begin{align}
{\cal S}[B]  &= - \frac{1}{kg^2}  {\rm Tr} \left\{ \frac{1}{2} \sum_{I=1}^2 (  [\phi,v_I][\bar{\phi},v_I] - \psi_I [ \bar{\phi} , \psi_I ]  + \psi_I [ v_I , \eta ] )
 - \frac{1}{8}[\phi,\bar{\phi}]^2  \right.\\
&\left. - B^2 + i {\cal E} B + \frac{1}{8} \eta [ \phi , \eta] + \chi [ \phi , \chi ] + \sqrt{2} 
 \chi \left(  [v_1 , \psi_2 ] + [\psi_1 , v_2 ] \right)
 \right\}~.
 \end{align}
 Here ${\cal E}= \sqrt{2} i [v_1,v_2]$. 
A BRST-like charge ${\cal Q}$ is defined as 
$
{\cal Q} \phi =0,
{\cal Q}\bar{\phi} = \xi~,
{\cal Q} \xi = [\phi , \bar{\phi} ]~,
{\cal Q} v_I = \psi_I ~,
{\cal Q} \psi_I = [\phi , v_I ]~,
{\cal Q} \chi = B~,
{\cal Q} B = [ \phi , \chi ]~,
$
where ${\cal Q}$ is nilpotent up to traceless part: 
${\cal Q}^2 \phi = 0 ~, Q^2 \bar{\phi} =[\phi , \bar{\phi} ],
Q^2 v_I = [ \phi , v_I ]~~,\mbox{etc} $
($I=1,2$). 
The action ${\cal S}[B]$ is written in the ${\cal Q}$ closed form, 
\begin{align}
{\cal S}[B]&=   Q {\cal P}~,&
{\cal P} &= \frac{1}{kg^2}  \left\{ \frac{1}{2}  \sum_{I=1}^2 \psi_I [v_I , \bar{\phi}] 
 + \frac{1}{8}  \eta [\phi,\bar{\phi}] + \chi  B - i  \chi {\cal E}  \right\}~.
\end{align}
Thus, ${\cal S}[B]$ is ${\cal Q}$-exact. 
The partition function is not a path integral, but an ordinary integration,
\begin{align}
{\cal Z} &= \frac{1}{{\rm vol}(G)} \int [d v][d \lambda ][dB] \exp ( i Q {\cal P} ) ~.
\end{align}
All fermionic contents form a finite-dimensional Grassmann algebra which are generated by $\{ \Lambda_{\alpha}^a \}$.  The Grassmann algebra are graded and we call the number of multiplied $\Lambda_{\alpha}^a$ rank.  
Let us understand that the fermionic integration pick up the highest rank elements. 
This integral might be divergent because there are flat directions, though we take account into the Euclideanization of $\bar{\phi}$ and there is the vanishing determinant term which comes from the fermionic part.  
Let us change our space to Euclidean space $-{\cal S}[B]_{\rm E} := i {\cal S}[B]$. 
\begin{align}
{\cal Z}_{\rm E} &= \frac{1}{{\rm vol}(G)} \int [d \phi][d\bar{\phi}][dB ][dv_I][d \eta] [d\chi] [d \psi_I ] \exp ( - Q {\cal P} )= \frac{1}{{\rm vol}(G)} \int [d \phi] \exp( - {\cal S}_{\rm eff} )~.
\end{align}

Let us assign ghost charge $+1$ to $Q$. The Lagrangian should have ghost charge $0$.  
If the ghost charge of $\chi$ is $\nu$, the auxiliary field $B$ has ghost charge $\nu+1$. Then the term $Q {\rm Tr} \chi B$ should have charge $0$ and it means that $1+\nu+(\nu+1)=0$. So $\nu=-1$. The matrix $\phi$ have the same charge as that of $Q^2$. The charge of $\phi$ is $2$ and $\bar{\phi}$ has $-2$. From the term $Q {\rm Tr} \chi {\cal E}$, we can read off the charge of $v_I$ as $0$.
\begin{table}
\begin{tabular}{cccccccc}
$Q$ & $\phi$ & $\bar{\phi}$ & $\eta$ & $\chi$ & $B$ & $v_I$ & $\psi_I$\\
\hline
$1$ & $2$ & $-2$ & $-1$ & $-1$ & $0$ & $0$ & $1$ 
\end{tabular}
\caption{The assignment of the ghost charge.}
\end{table}
Let us deform the BRST charge with respect to the little group SO(2):
${\cal Q} \rightarrow {\cal Q}_{\varepsilon}$,
\begin{align*}
{\cal Q}_{\varepsilon} v_I &= \psi_I ~, 
&{\cal Q}_{\varepsilon}\bar{\phi} &= \xi~,
&{\cal Q}_{\varepsilon} \chi &= B~,
&{\cal Q}_{\varepsilon} \phi &=0~,\\
{\cal Q}_{\varepsilon} \psi_I &= [\phi , v_I ] + \varepsilon_{IJ} E v_J~,
&{\cal Q}_{\varepsilon} \xi &= [\phi , \bar{\phi} ]~,
&{\cal Q}_{\varepsilon} B &= [ \phi , \chi ] + E \chi~.
\end{align*}
Here $E$ is the deformation parameter and let us consider the deformed action:  
${\cal S}_{\varepsilon} = {\cal Q}_{\varepsilon} {\cal P}$. 
Let us assign the ghost charge of $E$ as $0$. Then the additional terms coming from this deformation have ghost charge $-1$ and it does not matter. Therefore we can consider this deformation does not change the value of integration. 

Localization technique shows that the integral reduces to counting of the fixed points.   
The integral is Gaussian for almost all fermions and $B$ and $\bar{\phi}$.  
We add mass terms for the convergence of zero-modes' integrations by hand. After that
in the classical limit (this procedure picks up the fixed point),
 we can integrate them and obtain the simple result
\begin{align*}
{\cal Z} &= \frac{1}{{\rm vol}(G)}\int [d \phi] \frac{1}{{\rm Det}_{{\rm adj.}}({\rm ad} \phi + E)}
\end{align*}
where we neglect various numerical factors coming from the Gaussian integrations. We expect that such a numerical factor cancels for pairs of boson and fermion except for $\phi$. 
In addition, though the first integral converges, the resultant integral diverges. It means that there exists a space to discuss the validity of the deformation method. 
Let us consider this integral with some regularization, which may not be equal to the original integral.
The integrand is invariant under a transformation $\phi \rightarrow \phi' = U \phi U^{-1}$ with a unitary matrix $U$. 
Let us use the Weyl integration formula. These integrals reduce to integrals over the Cartan subalgebra.
The volume of group is given as ${\rm vol}({\rm SU}(N)/{\mathbb Z}_N)$.
The integral can be reduced to a simple integral over diagonal traceless matrices. 
\begin{align*}
{\cal Z} [A_{N-1}] &= \frac{N}{ E^{N-1} N!} \oint [d \phi] \prod_{ i < j} \frac{(\phi_i - \phi_j)^2}{(\phi_i - \phi_j)^2 - E^2}~,
\end{align*}
where $N$ in the numerator is order of the center of SU($N$) and 
$N!$ is the order of its Weyl group. 
Generalizations to other groups are obtained immediately. Let $\Phi$ be one of root systems $A_{N-1} , B_N , C_N , D_N$. The integral for $\Phi$ is deduced as
\begin{align*}
{\cal Z} [\Phi] &= \frac{\# Z}{ E^{r} \# W} \oint [d \phi] \prod_{ \alpha \in \Phi} \frac{( \alpha , \phi )}{( \alpha , \phi ) - E} ~.
\end{align*}
Here in order to obtain finite result, we close the contour with additional paths. We will talk the detail later. 
The symbol $\alpha$ denotes a root, the variable $\phi$ is an element of CSA, $Z$ is the center of the universal covering group of $G$, and $W$ is the Weyl group. 
This integral diverges because poles sitting on the contour. In order to avoid such a divergence, let us shift the parameter $E$ to the imaginary direction ${\rm Im}(E) > 0$. 
Then we count poles where $(\alpha , \phi) > 0$ for all positive roots $\alpha$ because we chose the upper contour on each complex $\phi_i$-plane. This region is a Weyl chamber. 
Let us define quantities $p_i = ( s_i , \phi)$ where $s_i (i=1, \cdots , r)$  are simple roots. 
The linear transformation $\phi \rightarrow (p_i)$  yields a Jacobian $1/C$ where $C$ is the determinant of the Cartan matrix and 
$C=\# Z$. 
Our previous work shows that each nontrivial pole which exists  inside of a Weyl chamber is given by $p_i=E$ for all $i$ for all classical gauge groups, 
where $p_i$ are defined by the fundamental root system obtained by the Weyl chamber.  
Let us put ourselves on a skeptical ground. In this article we do not treat the result by MNS as exact.
Our selection of contour excludes the additional multiplicity. This is not a truncation. This differs from that of MNS. This contour is, in a sense, a fundamental cycle and MNS count $(N-1)!$ cycles.  
Remaining poles are sitting on boundaries of Weyl chambers. 
The consideration on poles on boundaries are more sensitive.
In this paper we ignore such a pole on boundaries. 
Such a poles might exist in SO and USp cases, while in SU model, there are no poles on boundaries. 
Concrete calculation show that the residue at poles on boundaries in some cases in SO and USp models vanish. 
This is the reason why we do not consider the contribution of poles on boundaries \kcor{\cite{KMLee}}.
In addition, diagrams which are not connected give points on boundaries. 

Every root $\alpha$ is written as a linear sum of simple roots with integral coefficients and this fact inherits to the variable $(\alpha , \phi)$. 
\begin{align}
\alpha &= \sum_{i=1}^r u(\alpha)_i s_i~,& (\alpha , \phi) &= \sum_{i=1}^r u(\alpha)_i p_i~.
\end{align}
 Therefore the integral is written as 
\begin{align*}
{\cal Z} [\Phi] &= \frac{1}{ \# W}  \left( \prod_{ \alpha \in \Phi^+, \alpha \notin \{ \alpha_j\} } \prod_{i=1}^r \frac{u(\alpha)_i}{u( \alpha)_i - 1} \right) \left( \prod_{\alpha \in \Phi^+} \frac{u(\alpha)_i}{u( \alpha)_i + 1} \right)~.
\end{align*}
In order to obtain the result $1/N^2$, we need multiplicity $(N-1)!$ for SU($N$) cases. We do not insert the multiplicity in this paper. We show several value for low rank groups in Table \ref{tbl:lowrk}.
\begin{table}
\begin{tabular}{|c|c||c|c||c|c||c|c|}
\hline
$A_1$ & $C_1$ & $A_1 \oplus A_1$ & $D_2$ & $B_2$ & $C_2$  &$A_3$ & $D_3$ \\
\hline
$1/4$ & $1/4$ & $1/4 \times 1/4$ & $1/16$ & $3/64$ & $3/64$ & $1/96$ & $1/96$  \\
\hline
\end{tabular}
\caption{This table show that the resulting residues correspond with the isomorphic algebra.}
\label{tbl:lowrk}
\end{table}
These poles are written as 
\begin{align}
A_{N-1}&: \phi_i = (N+1)/2-i,& 
B_{N}&: \phi_i= N-i+1,&
C_N&: \phi_i = N-i+1/2,&
D_N&: \phi_i = N-i,
\end{align}
where $1 \leq i \leq N$. We can interpret these poles as a configuration of branes.  Residues are 
computed for all classical groups. 
\begin{align}
{\cal Z}[A_{N-1}] &= \frac{1}{N \cdot N!}~,&
{\cal Z}[D_{N}] &= \frac{1}{2^{N}N \cdot N!} \frac{(2N-3)!!}{(2N-4)!!}~,&
{\cal Z}[B_{N}] &= {\cal Z}[C_{N}] = \frac{1}{2^{N+1} N \cdot  N!} 
\frac{(2N-1)!!}{(2N-2)!!}~,
\end{align}
Equivalence relations, ${\cal Z}[B_N] = {\cal Z}[C_N]$ and $(N+1)^2 {\cal Z}[D_{N+1}] =  N {\cal Z}[C_N]$, are derived. 
\subsection{Grand Partition Functions; A,C} 
Now we can sum up them. 
Let us define corresponding grand partition functions;  
\begin{align}
\Theta[A;q] &:= 1+ \sum_{N=2}^{\infty} {\cal Z}[A_{N-1}] q^N = 1+ \sum_{N=2}^{\infty} \frac{1}{N} \frac{q^N}{N!}\\
\Theta[D;q] &:= 1  + \sum_{N=2}^{\infty} {\cal Z}[D_N] q^{2N} = 1 + \sum_{N=2}^{\infty}
 \frac{1}{N} \frac{(2N-3)!!}{(2N-4)!!} \frac{1}{N!}\left( \frac{q^2}{2} \right)^N~,\\
\Theta[B;q] &:=q+ \sum_{N=1}^{\infty} {\cal Z}[B_N]q^{2N+1} =q+ {q} \sum_{N=1}^{\infty} 
\frac{(2N-1)!!}{(2N)!!} \frac{1}{N!} \left( \frac{q^2}{2} \right)^N\\
\Theta[C;q] &:=1+ \sum_{N=1}^{\infty} {\cal Z}[B_N]q^{2N} =1+ \sum_{N=1}^{\infty} 
\frac{(2N-1)!!}{(2N)!!} \frac{1}{N!} \left( \frac{q^2}{2} \right)^N
\end{align}
where $q=e^{-\beta}$ and $\beta$ is the chemical potential which couples to the matrix size $N$. 
The convergent radii of these functions are infinite.   
These functions satisfy the relation $\Theta[B;q]=q\Theta[C;q]$ and written in terms of the exponential integral function ${\rm Ei}(z)$ and the confluent hypergeometric function $\Psi(s , t; z)$ defined in the appendix. Here we study about grand partition functions of $A$ and $C$ series;
\begin{align}
\Theta[A;q]_+ &= 1-q+ \int_0^q \frac{e^x-1}{x} dx =
{\rm Ei}(q)-{\bf C} - \ln(q) - q +1~,&
\Theta[C;q]_+ &= \Phi(1/2,1;q^2/2)~,
\end{align}
These functions, $\Theta[A;q]_+$ and $\Theta[C;q]_+$, obey  the following differential equations. 
The meaning of the subscript $+$ is explained later.  
\begin{align}
\left[ \frac{d^2}{dq^2} -(1-q^{-1}) \frac{d}{dq}  \right] \Theta[A;q]_+ &= 1~,&
 \left[ \frac{d^2}{dq^2} -  (q-q^{-1}) \frac{d}{dq} - 1   \right] \Theta[C;q]_+ &=0
\end{align}
These equations are second order linear differential equation with polynomial coefficients. 
The equation of $\Theta[A;q]_+$ is inhomogeneous while the equation of $\Theta[C;q]_+$ is homogeneous. 
Our series solutions are particular solutions of these equations. 
The corresponding homogeneous equations are
 \begin{align}
\left[ \frac{d^2}{dq^2} -(1-q^{-1}) \frac{d}{dq}  \right] \Theta[A;q]_+ &= 0~,&
 \left[ \frac{d^2}{dq^2} -  (q-q^{-1}) \frac{d}{dq} - 1   \right] \Theta[C;q]_+ &=0
\end{align}
These equations have one singular point $q=0$ of first kind and one $q=\infty$ of second kind. 
Our solutions are regular at $q=0$. Because the singular point $q=0$ is first kind, we can consider their indicial equations.
\begin{align}
\nu (\nu -1) + \nu &=0~,&
\mu (\mu -1) + \mu &=0~.
\end{align}
This means that general solutions of these equations might have $\log$ divergence at $q=0$. 
Polynomial approximation is not so bad because the convergent radii are infinite. However if we consider their logarithm, sometimes we are led to wrong direction.  
In order to avoid it, we use differential equation though its validity in $q>>1$ region is not so clear. 

Connected generating function ${\cal W}$ is log of grand partition function $\Theta$; ${\cal W} = -\ln \Theta$ in Euclidean model. The expectation value of the matrix size $N$ are defined as the derivative of ${\cal W}[O;\beta]:= -\ln \Theta[O; q]$ where the symbol $O$ represents the type of sequence $O=A,C$;  $\varphi = \langle N \rangle_{O}= (\partial/\partial \beta) {\cal W}[O;\beta]$. 
The effective action $\Gamma[O;\varphi]$ is defined by the Legendre transformation; $\Gamma[O;\varphi] := {\cal W}[O;\beta] - \beta \varphi$. We will show their graph later. 
\begin{figure}[htb]
\psfrag{beta}{\huge $\boldsymbol{\beta}$}
\psfrag{<N>}{\huge $\boldsymbol{\langle N \rangle}$}
\begin{tabular}{cc}
\psfrag{Theta}{{\huge $\boldsymbol{\Theta}$}}
\psfrag{Th[A]+}{{\huge $\boldsymbol{\Theta[A]_+}$}}
\psfrag{Th[A]-}{{\huge $\boldsymbol{{\Theta}[A]_-}$}}
\psfrag{Th[C]+}{{\huge $\boldsymbol{\Theta[C]_+}$}}
\psfrag{Th[C]-}{{\huge $\boldsymbol{{\Theta}[C]_-}$}}
\rotatebox{270}{\resizebox{50mm}{!}{\includegraphics{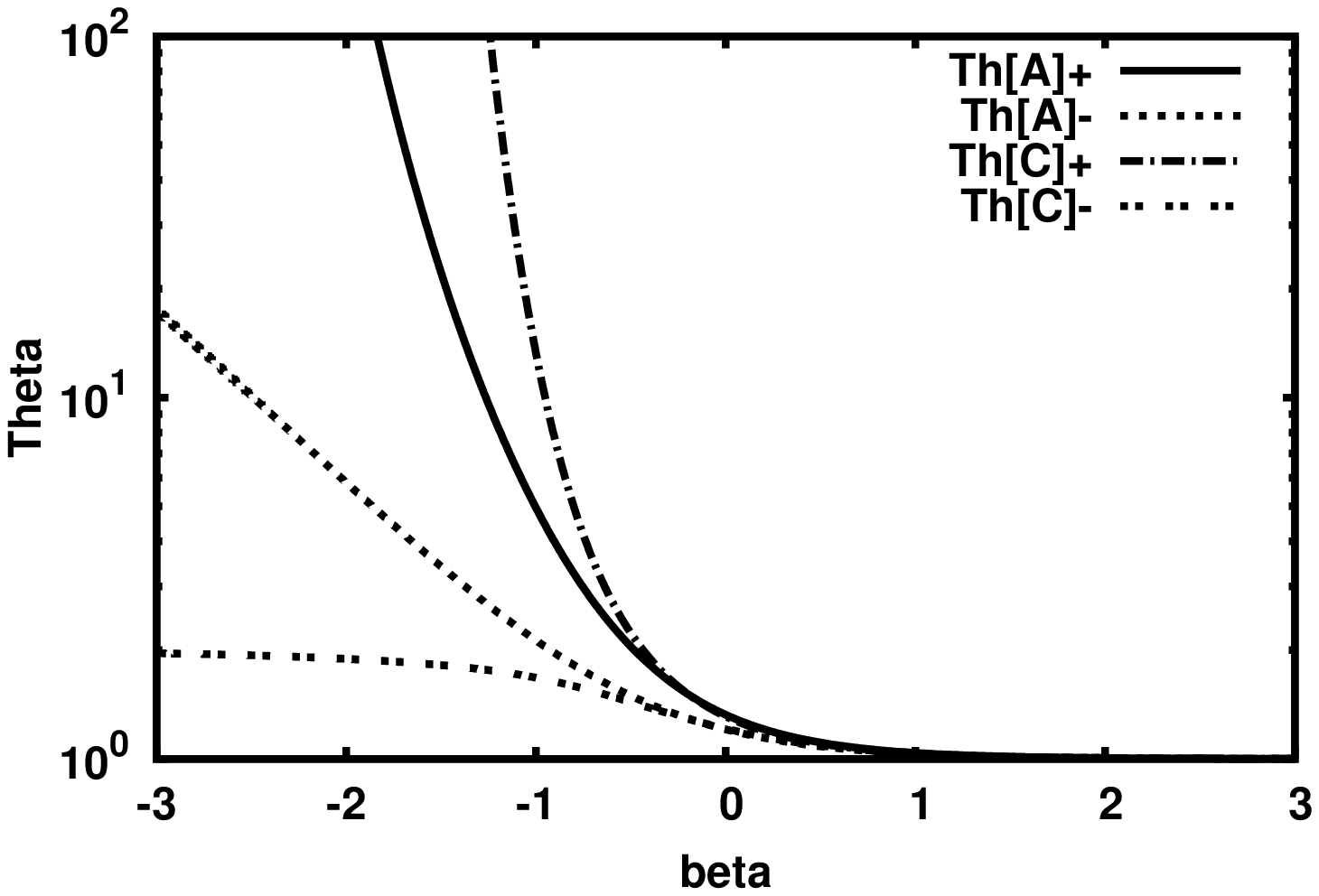}}}
&
\psfrag{<N>[A]+}{\huge $\boldsymbol{\langle N \rangle_{A+}}$}
\psfrag{<N>[A]-}{\huge $\boldsymbol{\langle N \rangle_{A-}}$}
\psfrag{<N>[C]+}{\huge $\boldsymbol{\langle 2N \rangle}_{C+}$}
\psfrag{<N>[C]-}{\huge $\boldsymbol{\langle 2N \rangle_{C-}}$}
\rotatebox{270}{\resizebox{50mm}{!}{\includegraphics{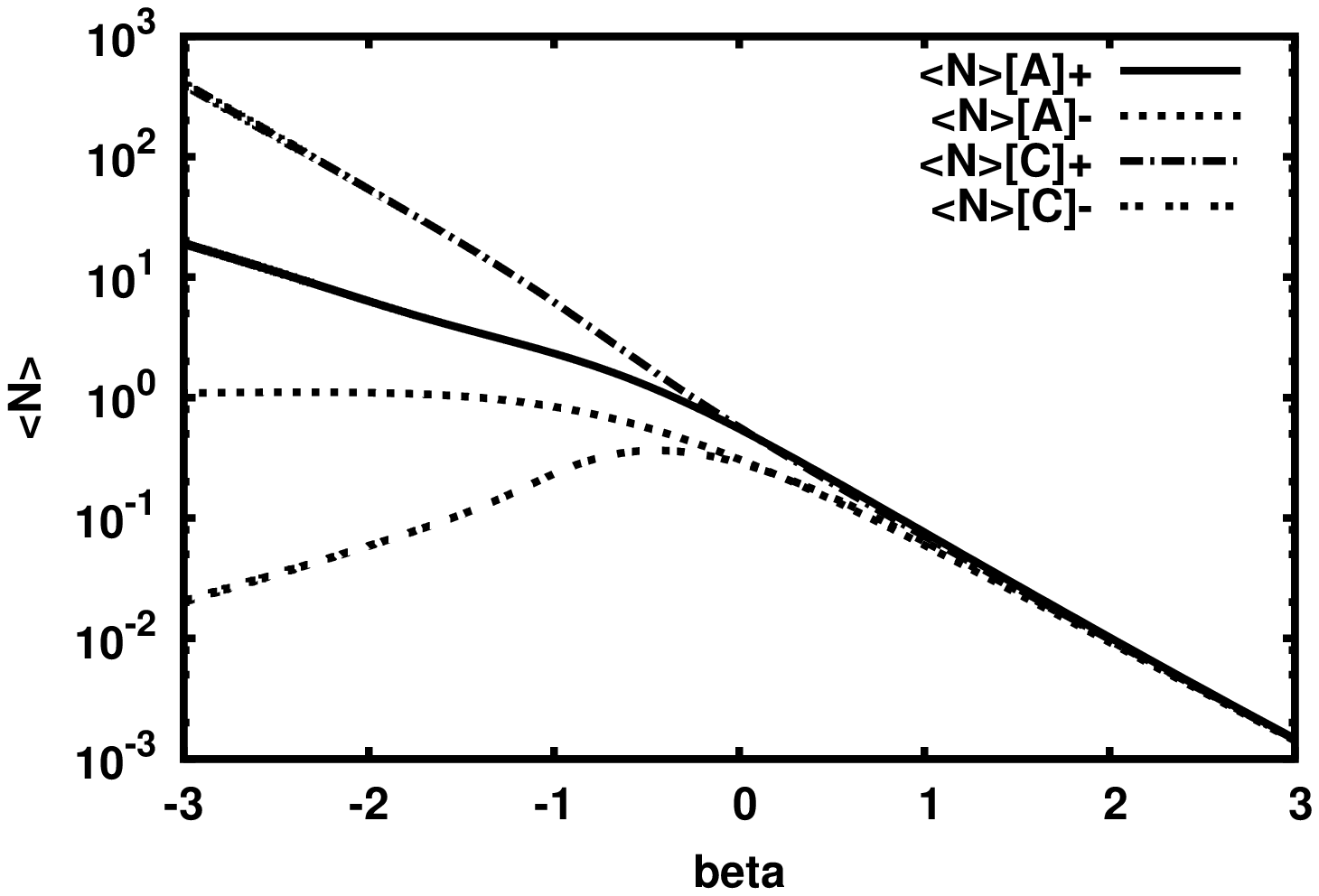}}}
\end{tabular}
\caption{Grand partition functions and the expectation values of the matrix size $N$ and 2$N$ for A and C, respectively. }
\label{fig:gpf}
\end{figure}
\begin{figure}[htb]
\psfrag{<N>}{\huge $\boldsymbol{\langle N \rangle}$}
\psfrag{Gam}{\huge $\boldsymbol{\Gamma}$}
\psfrag{Gam[A]}{\huge $\boldsymbol{\Gamma[A]_+}$}
\psfrag{Gam[A]m}{\huge $\boldsymbol{\Gamma[A]_-}$}
\psfrag{Gam[C]}{\huge $\boldsymbol{\Gamma[C]_+}$}
\psfrag{Gam[C]m}{\huge $\boldsymbol{\Gamma[C]_-}$}
\begin{tabular}{cc}
\rotatebox{270}{\resizebox{50mm}{!}{\includegraphics{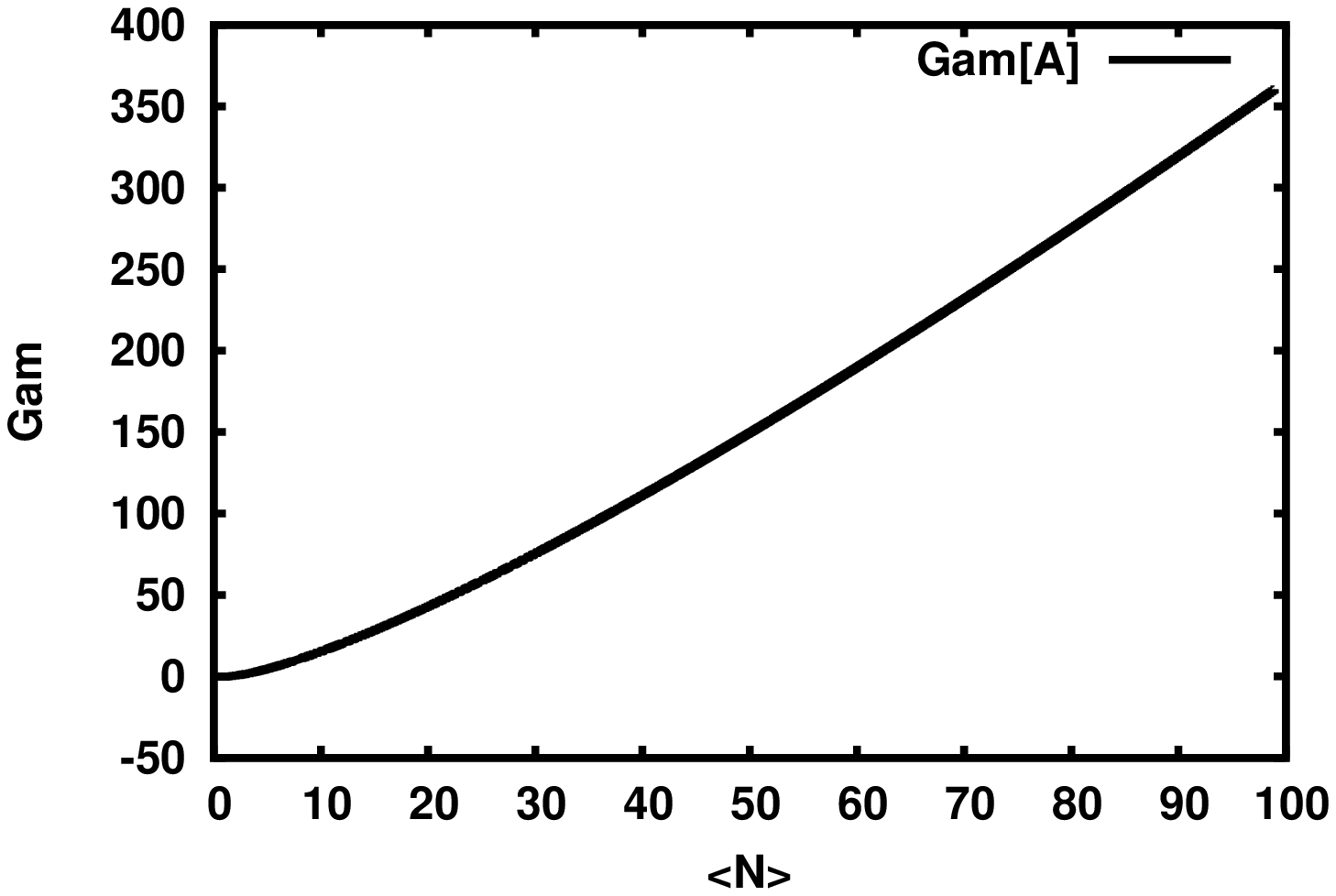}}}&
\rotatebox{270}{\resizebox{50mm}{!}{\includegraphics{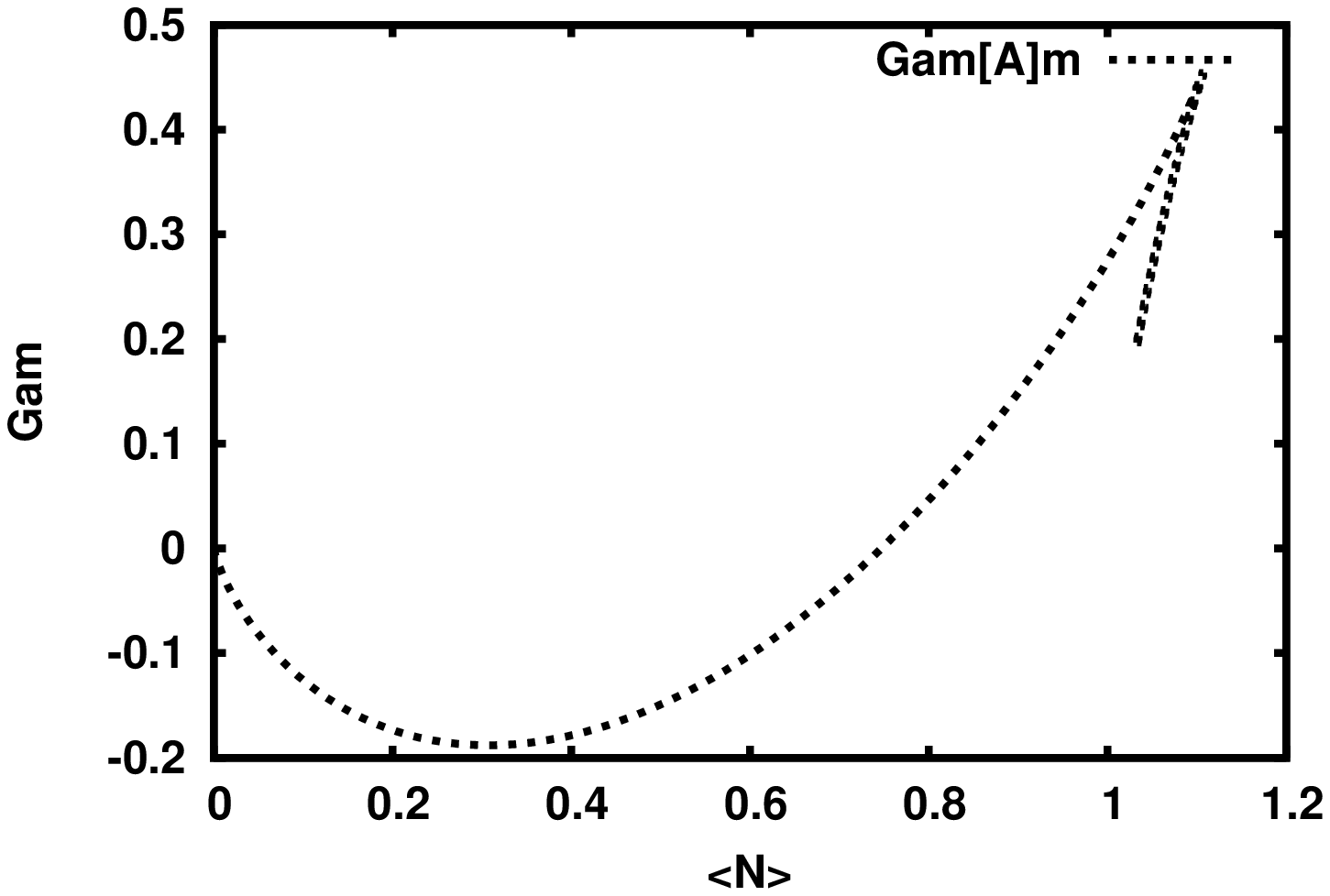}}}\\
\rotatebox{270}{\resizebox{50mm}{!}{\includegraphics{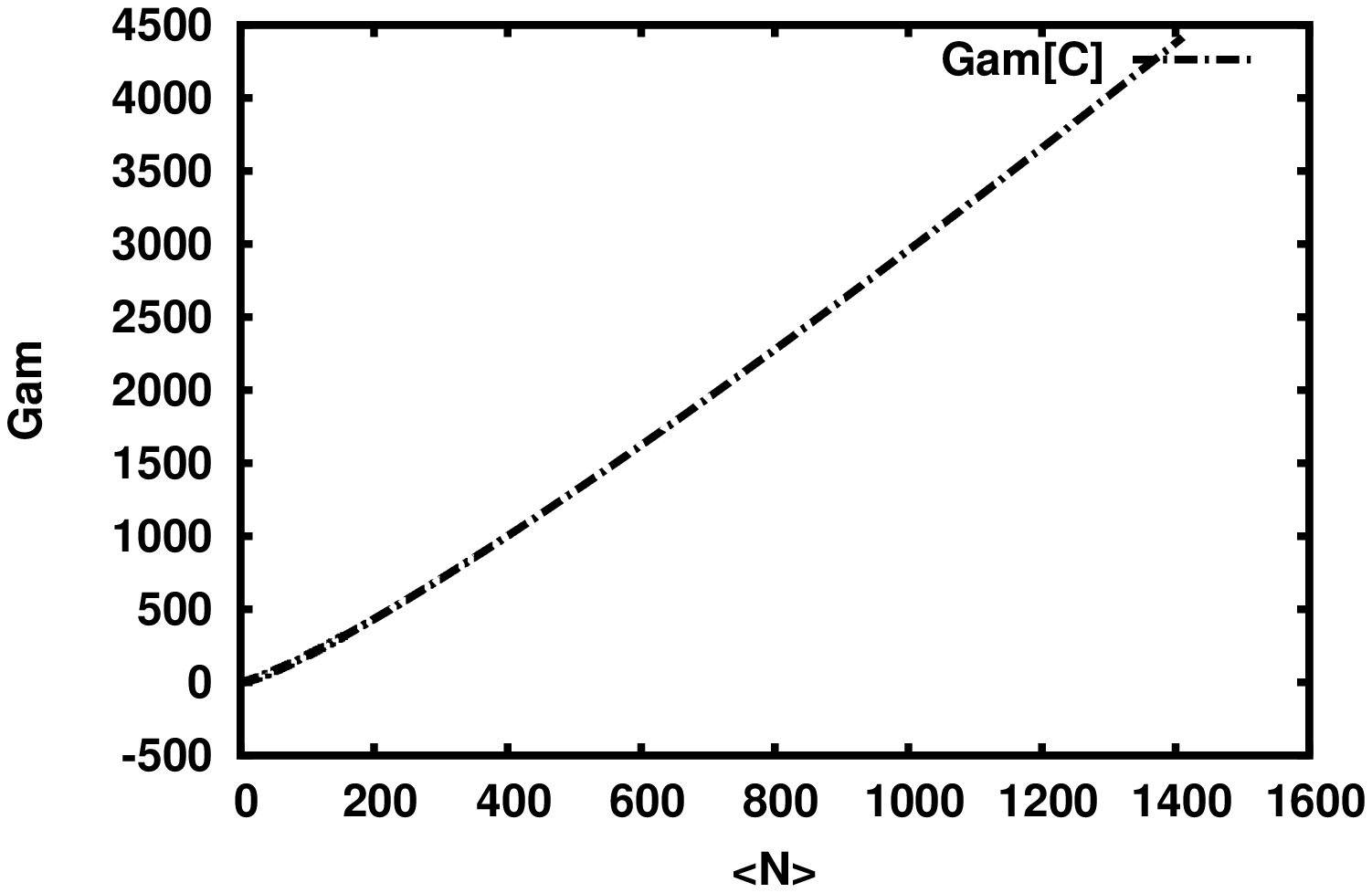}}}&
\rotatebox{270}{\resizebox{50mm}{!}{\includegraphics{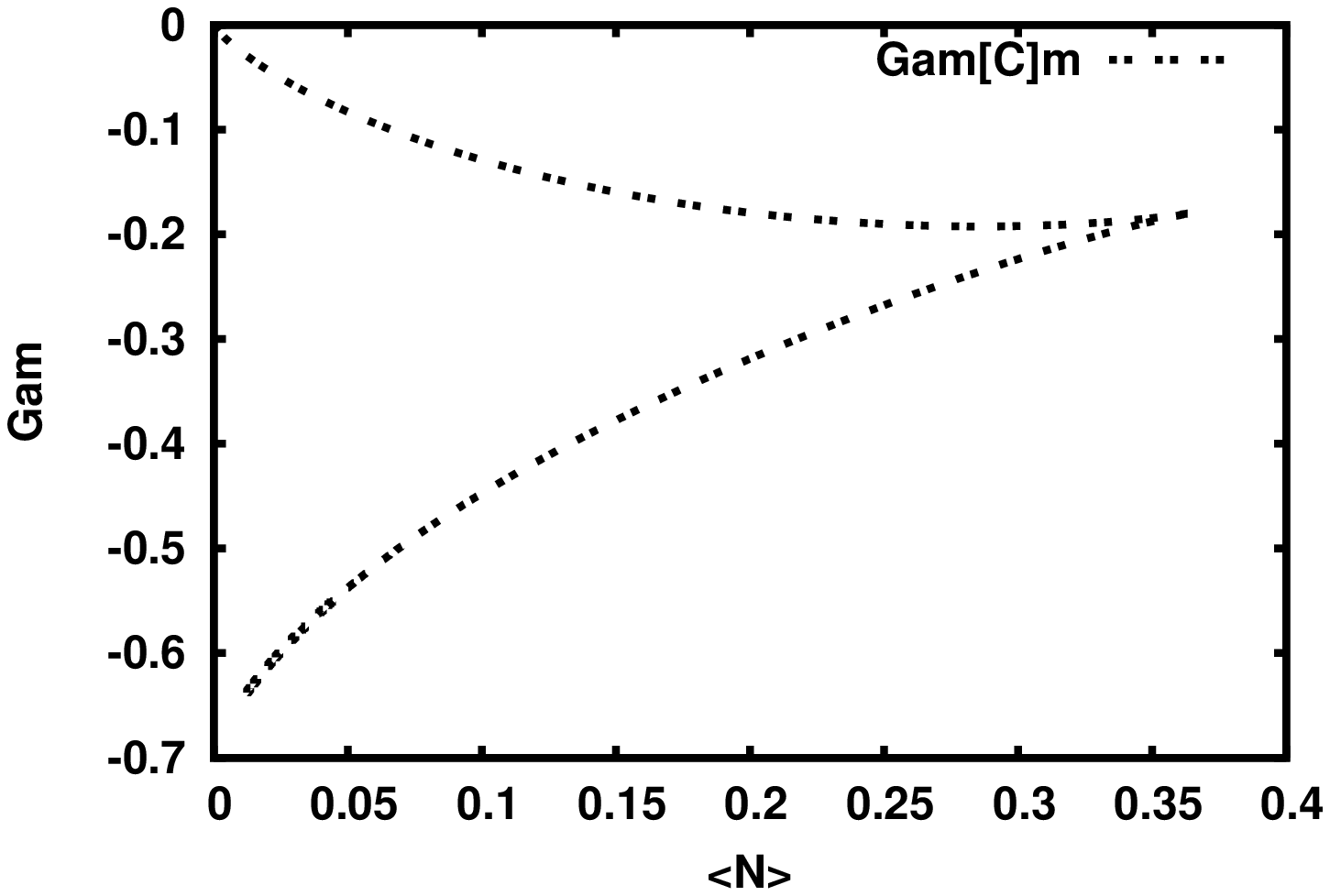}}}
\end{tabular}
\caption{Effective potentials $\Gamma[A]_{\pm}$, $\Gamma[A]_{\pm}$ as functions of the expectation value of matrix size, $\langle N \rangle$.}
\end{figure}
In the evaluation of matrix integrals, we omit the discussion on sign. 
Let us consider grand partition functions with sign; 
\begin{align}
{\Theta}[A;q]_- &:= 1+ \sum_{N=2}^{\infty} {\cal Z}[A_{N-1}] (-q)^N&
{\Theta}[C;q]_- &:=1+ \sum_{N=1}^{\infty} {\cal Z}[B_N](-)^{N-1}q^{2N} ~.
\end{align}
We use the subscript $\pm$ in order to explain this difference.   
They follow to the following differential equations;
\begin{align}
\left[ \frac{d^2}{dq^2} +(1+q^{-1}) \frac{d}{dq}  \right] {\Theta}[A;q]_- &= 1~,&
 \left[ \frac{d^2}{dq^2} +  (q+q^{-1}) \frac{d}{dq} + 1   \right] {\Theta}[C;q]_- &=2~.
\end{align}
Their graphs are included to the previous figure and we show their effective potential. 
The graphs for $\Theta_-$ have cusp singularities. 
These effective potentials which are functions of the expectation value of the size of matrix are not single valued.  There are two branches in SU and USp cases.
These results imply the phase transition of these models {with respect to the variation of the expectation value of the size of matrix}.

\section{Conclusion}
We reviewed non-Abelian Berry's phases in the USp matrix model as an effect of fermionic integration against the classical background. We encountered generalized monopoles. 
As such a monopole, we studied about the Tchrakian's five dimensional monopole. The charges of the Tchrakian's monopole and the Yang monopole are instanton number. 
The effect of Berry's phase is considered in the T-dualized model. 
In order to interpret them in terms of the original model, we performed matrix integrals. 
We obtain full result for all of classical gauge groups and their grand partition functions are computed. The grand partition functions are written as an exponential integral function and confluent hypergeometric functions. 
We showed graphs of grand partition functions of A and C series. In other words, matrix integral for the little IIB and little USp matrix model were exhibited.  
In these models, the expectation values of numbers $N$ and their effective potentials were considered. 
There exist a problem of choice of sign. 
In the case of minus sign, we obtain cusp singularities. 
It shows that there are another phases of these model. 
From the figure the start points which correspond with $q=0$ are not stable. If we assume that these models have sense only when the matrix sizes are large enough, the reliable regions are around $q=1$ and they are neighbor of those cusps.  
 The graph for $SU(N)$ show that the potential fall into the expectation value $<N>=1$, while for $USp(2N)$ the stable point looks like $<N>=0$.  
We would like to compute matrix integral for full Itoyama-Tokura model.  
There are several interesting papers which might be related to our work \cite{Pedder:2008je}.

\section*{acknowledgment}
This work has been supported by Osaka University,
 JSPS(01J00801), COE program in Osaka City University(Constitution of wide-angle mathematical basis focused on knots), Korea Institute for Advanced Study. 
I would like to thank to Hiroshi Itoyama, Asato Tsuchiya, Reiji Yoshioka, Leonard Susskind, Ki-Myeong Lee, Piljin Yi for various advices.  
\kcor{I would like to thank to Jun Nishimura for his seminar and advices. }
I would like to thank to all of my colleague in Osaka University, Osaka City University, KIAS and all visitors.  

\appendix
\section{Basic Tables on Classical Lie groups}
Let us list some of linear groups. We omit the Lorentz groups. 
\begin{align}
GL(N,K) &:= \{ g \in M(N,K)| \det g \neq 0 \}~,&
SL(N,K) &:= \{ g \in M(N,K)| \det g = 1 \}~,\cr
U(N) &:= \{ g \in M(N,{\mathbb C}) | g^{\dag} = g^{-1}  \}~,&
SU(N) &:= U(N) \cap SL(N,{\mathbb C})~,\cr
O(N) &:= \{ g \in M(N,{\mathbb R}) | g^t = g^{-1} \}~,&
SO(N) &:= O(N) \cap SL(N,{\mathbb R})~,\cr
USp(2N)&:= \{ g \in U(2N) | g^tJ_{2N} g=J_{2N}  \}~,&
J_{2N} &= \begin{pmatrix} 0 & {\bf 1}_N \\ - {\bf 1}_N & 0 \end{pmatrix} ~,
\end{align}
where $K= {\mathbb R}$ or ${\mathbb C}$ and $N \geq 1$.  
Group $O(1)$ is a discrete group which is isomorphic to $\{ \pm 1 \}$.  
Group $SU(1)$ is a unit group $\{ 1 \}$. 
Groups $U(1),SO(2)$ are isomorphic to each other. They are commutative and their fundamental groups are $\pi_1(U(1))  \simeq \pi_1(SO(2)) \simeq {\mathbb Z}$. 
We will omit these groups. 
Group $SO(N) (N \geq 3)$ is not simply connected, but doubly connected;  $\pi_1(SO(N)) \simeq {\mathbb Z}_2 (N \geq 3)$. 
The universal covering group of $SO(N)$ is denoted by $Spin(N)$. 
Groups $SU(N) (N \geq 2)$ are simply connected. 
group $Spin(N)$ is formed in terms of Clifford algebra. 
Let us list their centers. We denote the center of a group $G$ $Z(G)$. The Schur's lemma show that the center of the general linear group $GL(N,{\mathbb C})$ is the group which consists of scalar operations $\lambda {\bf 1}_N (\lambda \in {\mathbb C}^{\times})$ where $k^{\times} = k \setminus \{ 0 \}$.   
\begin{align}
Z(GL(N,K)) &\simeq K^{\times}~,& Z(SL(N,K)) &\simeq \{ z \in K^{\times} , z^N =1 \}~,\cr
Z(U(N)) & \simeq S^1\simeq \{ z \in {\mathbb C}^{\times} | |z| = 1 \}~,& Z(SU(N)) &\simeq {\mathbb Z}_N \simeq \{ z \in {\mathbb C}^{\times} | z^N =1  \} ~,\cr
Z(SO(2N)) &\simeq  {\mathbb Z}_2~, &Z(SO(2N+1)) & \simeq 1~,\cr
Z(Spin(4N)) & \simeq {\mathbb Z}_2 \times {\mathbb Z}_2~,&
Z(Spin(4N+2)) & \simeq {\mathbb Z}_4~,\cr
Z(Spin(2N+1)) & \simeq {\mathbb Z}_2~,&
Z(USp(2N)) &\simeq {\mathbb Z}_2~,
\end{align}
There are exact sequences; $ 1 \rightarrow \pi_1(SO(N)) \rightarrow Z(Spin(N)) \rightarrow Z(SO(N)) \rightarrow 1 \mbox{~(exact)}$. 
Their Lie algebras are
\begin{align}
gl(N,K) &:=  M(N,K)~,&
sl(N,K) &:= \{ X \in M(N,K)| {\rm Tr} X = 0 \}~,\cr
u(N) &:= \{ X \in M(N,{\mathbb C}) | X^{\dag} + X = 0  \}~,&
su(N) &:= u(N) \cap sl(N,{\mathbb C})~,\cr
so(N) &:= \{ X \in M(N,{\mathbb R}) | X^t + X = 0 \}~,&
usp(2N)&:= \{ X \in u(2N) | X^tJ_{2N} +J_{2N}X = 0  \}~,
\end{align}
where $gl(N,{\mathbb C}), sl(N,{\mathbb C})$ are complex Lie algebra and the remaining are real Lie algebra. The complexification of $su(N)$ is $sl(N,{\mathbb C})$. We denote the complexification of Lie algebra $so(N),usp(2N)$ $so(N)^{\mathbb C}, usp(2N)^{\mathbb C}$, respectively. Complex Lie algebra $sl(N,{\mathbb C}) (N \geq 2), so(N)^{\mathbb C} (N \geq 3, N \neq 4), usp(2N)^{\mathbb C} (N \geq 1)$ are simple. There are several isomorphisms of Lie algebra; $so(3) \simeq su(2)\simeq usp(2)$, $so(5) \simeq usp(4)$ and $so(4) \simeq su(6)$. 

Let us focus on Lie algebra $sl(N,{\mathbb C}) (N \geq 2), so(N)^{\mathbb C} (N \geq 3, N \neq 4), usp(2N)^{\mathbb C} (N \geq 1)$. 
For shortening our notation, we use symbol $sl_N, so_N,usp_{2N}$ for them respectively. 
Let us consider their Cartan decomposition. For each algebra ${\mathfrak g}$, ${\mathfrak h}({\mathfrak g})$ denotes a Cartan subalgebra of ${\mathfrak g}$. 
\begin{align}
{\mathfrak h}(sl_N) &= \left\{ {\rm diag}( \phi_1 , \phi_2 , \cdots ,  \phi_N ) ,  \phi_1 + \cdots + \phi_N = 0 \right\}~,&
{\mathfrak h}(usp_{2N}) &= \left\{ \begin{pmatrix} & \Lambda \\ -\Lambda & \end{pmatrix} ~,
\Lambda = {\rm diag} ( \phi_1 , \phi_2 , \cdots ,  \phi_N ) \right\}~,\\
{\mathfrak h}(so_{2N}) &= \left\{ \begin{pmatrix}  & \phi_1 \\  -\phi_1 & \end{pmatrix} \oplus \cdots
\oplus \begin{pmatrix}  & \phi_N \\  -\phi_N & \end{pmatrix} \right\}~,&
{\mathfrak h}(so_{2N+1}) &= \left\{ \begin{pmatrix}  & \phi_1 \\  -\phi_1 & \end{pmatrix} \oplus \cdots
\oplus \begin{pmatrix}  & \phi_N \\  -\phi_N & \end{pmatrix} \oplus (0) \right\}~,
\end{align}
Of course, they are not unique. ${\mathfrak h}(usp_{2N})$ depends on the choice of $J_{2N}$.
 A Cartan subalgebra ${\mathfrak h}$  of a complex Lie algebra ${\mathfrak g}$ is a maximal toral subalgebra. 
The algebra ${\mathfrak h}$ acts on the algebra ${\mathfrak g}$ by commutator; for each $H \in {\mathfrak h}$ and for each $X \in {\mathfrak g}$ ${\rm ad} : {\mathfrak h} \rightarrow gl({\mathfrak g})$, ${\rm ad}(H)(X) = [H,X]$. 
The space ${\mathfrak g}$ is a complex vector space. Every irreducible representation of commutative algebra is one-dimensional. 
 Let $V$ be such a irreducible part of ${\mathfrak g}$. 
The dimension of $V$ is one and the action of ${\rm ad}(H)$ reduced to the multiplication of number $\alpha(H)$ where the operation ${\rm ad}(H)$ is linear in $H$ and these eigenvalues define a linear form on ${\mathfrak h}$; $ \alpha \in {\mathfrak h}^*$. 
Let us define a space $V_{\alpha}:= \{ X \in {\mathfrak g} | [H,X] = \alpha(H)X~, \mbox{ for each } H \in {\mathfrak h}   \}$ for each $\alpha \in {\mathfrak h}^*$.  
Let us define the root system with respect to this Cartan decomposition; $\Phi := \{ \alpha \in {\mathfrak h}^* | \dim_{\mathbb C} V_{\alpha} \neq 0, \alpha \neq 0  \}$. 
Therefore ${\mathfrak g} = {\mathfrak h} \oplus \sum_{\alpha \in \Phi} V_{\alpha}$. 
Because ${\mathfrak g}$ is a finite dimensional vector space, $\Phi$ is a finite set. 
Let us list the root systems for $sl_N,so_{2N+1},usp_{2N},so_{2N}$. 
\begin{enumerate}
\item { $\Phi(sl_N) = A_{N-1}$ case}\\
Sometimes the notation $A_N$ is used for the Lie algebra. We use this notation for the root system. 
Let us use the Cartan subalgebra of $sl_N$ which was introduced in the previous part; 
${\mathfrak h}(sl_N) = \left\{ {\rm diag}( \phi_1,  \cdots ,  \phi_N),  \phi_1 + \cdots + \phi_N = 0 \right\}$. 
Let us define linear forms $\alpha_{ij} ( i < j)$ such that $\alpha_{ij}(H) = \phi_i - \phi_j$ $(i <j, H \in {\mathfrak h}(sl_N))$.  They and their counterparts form the root system
$\Phi(sl_N) = A_{N-1} = \{ \pm \alpha_{ij} , ~ 1 \leq i < j \leq N ~ \}~.$

\item {$\Phi(so_{2N})= D_N$}\\
Let us use the Cartan subalgebra ${\mathfrak h}(so_{2N}) = \left\{ \phi_1 J \oplus \phi_2 J \oplus \cdots \phi_N J \right\}$, where $J = \sigma_2 = \begin{pmatrix} 0 & - {\bf i} \\ {\bf i} & 0 \end{pmatrix}$. 
Let us define linear forms $\alpha_{ij},\beta_{ij} ( i < j)$ such that $\alpha_{ij}(H) = \phi_i - \phi_j$ and $\beta_{ij}(H) = \phi_i + \phi_j$ $(i <j, H \in {\mathfrak h}(so_{2N+1}))$. 

The root system becomes $\Phi(so_{2N}) = D_{N} = \{ \pm \alpha_{ij} ,\pm \beta_{ij} ~ 1 \leq i < j \leq N ~ \}$. 
\item {$\Phi(so_{2N+1})= B_N$}\\
Let us use the Cartan subalgebra ${\mathfrak h}(so_{2N+1}) = \left\{ \phi_1 J \oplus \phi_2 J \oplus \cdots \phi_N J  \oplus (0) \right\}$. 
Let us define linear forms $\alpha_{ij}, \beta_{ij}, \gamma_k$ such that $\alpha_{ij}(H) = \phi_i - \phi_j$, $\beta_{ij}(H) = \phi_i + \phi_j, \gamma_k(H) = \phi_k$ $(1 \leq i<j\leq N, 1 \leq k \leq N , H \in {\mathfrak h}(so_{2N+1}) )$. 

The root system becomes $\Phi(so_{2N+1}) = B_{N} = \{ \pm \alpha_{ij} ,\pm \beta_{ij}, \pm \gamma_k ~ 1 \leq i < j \leq N, 1 \leq k \leq N ~ \}$. 

\item {$\Phi(usp_{2N})= C_N$}\\
Let us use the Cartan subalgebra 
$ \displaystyle
{\mathfrak h}(usp_{2N}) = \left\{ \begin{pmatrix} & {\bf i} \Lambda \\ - {\bf i}\Lambda & \end{pmatrix} ~,
\Lambda = {\rm diag} ( \phi_1 , \cdots , \phi_N)  \right\}$~. 
Let us define linear forms $\alpha_{ij}, \beta_{ij}, \gamma_k$ such that $\alpha_{ij}(H) = \phi_i - \phi_j$, $\beta_{ij}(H) = \phi_i + \phi_j, \gamma_k(H) = \phi_k$ $(1 \leq i<j\leq N, 1 \leq k \leq N , H \in {\mathfrak h}(so_{2N+1}) )$. 
The root system becomes $\Phi(usp_{2N}) = C_{N} = \{ \pm \alpha_{ij} ,\pm  \beta_{ij}, \pm 2 \gamma_k ~ 1 \leq i < j \leq N, 1 \leq k \leq N ~ \}$. 
\end{enumerate}

\subsection{Fundamental root system}
The following sets consist of positive roots with suitable ordering;
\begin{enumerate}
\item $A_{N-1}^+ = \{ \alpha_{ij} , ~ 1 \leq i < j \leq N ~ \}~.$
\item $B_{N}^+ = \{ \alpha_{ij} , \beta_{ij} , \gamma_k, ~ 1 \leq i < j \leq N, 1 \leq k \leq N ~ \}~.$
\item $C_{N}^+ = \{ \alpha_{ij} , \beta_{ij} , 2 \gamma_k ~ 1 \leq i < j \leq N, 1 \leq k \leq N ~ \}~.$
\item $D_{N}^+ = \{ \alpha_{ij} , \beta_{ij},  ~ 1 \leq i < j \leq N ~ \}~.$
\end{enumerate}
If we restrict the range of $\phi_i$ to real number, these roots define real linear forms. 
Therefore a root is denoted as a vector. 
For example $\alpha_{12}$ of $D_N$ can be expressed as $(+1,-1,0 , \cdots , 0)$. However in $A_N$ series, there are constraint $\phi_1 + \phi_2+ \cdots + \phi_N=0$. This implies that any linear form have ambiguity caused by $\kappa=(1,1,\cdots, 1)$. 
Let us consider a vector $(\phi_1, \cdots , \phi_N)$ which satisfies the constraint. 
Let $\alpha$ be a linear form which maps such a vector into real number. 
Then $\alpha + \lambda \kappa$ gives the same value as $\alpha$, where $\lambda$ is an arbitrary real number. We would like to use the representation whose component satisfy the relation $\alpha_1 + \cdots + \alpha_N = 0$. This constraint determines the ambiguous factor $\lambda$. It might be a kind of a gauge  slice. 

Now   we can define the lexicographic order, {\it i.e.} for each real vector ${\boldsymbol x} =(x_1 ,\cdots, x_N)$, ${\boldsymbol x} > 0$ if and only if $x_1 = 0 , \cdots, x_{k-1}=0$ and $x_{k}>0$ for some integer $k$. 
For two vectors ${\boldsymbol x}, {\boldsymbol y}$,  ${\boldsymbol x}> {\boldsymbol y}$ if and only if ${\boldsymbol x}- {\boldsymbol y}>0$. This order is linear order of ${\mathbb R}^N$ such that every two elements can be comparable. 
In this sense our positive roots $\alpha$ satisfy the relation $\alpha >0$. 
Let $\Phi$ be one of $A_N,B_N,C_N,D_N$ and $\Phi^+$ be the positive root system. 
Because the ordering is linear, there is the minimal element $s_1$. Again the set $\Phi^+ \setminus {\mathbb R} s_1$ has the minimal element. 
Let us find such a vector recursively, $s_{k+1} = \min\{ \Phi^+ \setminus \bigoplus_{i=1}^{k} {\mathbb R}s_i\}$ and we obtain a set $\{ s_1 , \cdots , s_N \}$. It is a basis of ${\mathbb R}^N$. 
\begin{enumerate}
\item $A_N$: $s_N = \alpha_{12}, s_{N-1} = \alpha_{23}, \cdots, s_1 = \alpha_{N,N+1}$.  
\item $B_N$: $s_N = \alpha_{12}, s_{N-1} = \alpha_{23} ,\cdots, s_{2} = \alpha_{N-1,N}, s_1 = \gamma_N$. 
\item $C_N$: $s_N = \alpha_{12}, s_{N-1} = \alpha_{23} ,\cdots, s_{2} = \alpha_{N-1,N}, s_1 = 2\gamma_N$.  
\item $D_N$: $s_N = \alpha_{12}, s_{N-1} = \alpha_{23} ,\cdots,s_{3} = \alpha_{N-2,N-1}, s_{2} = \beta_{N-1,N}, s_1 = \alpha_{N-1,N}$
\end{enumerate}
In fact, they are linearly independent. They form a lattice $L(\Phi)=\oplus_{i=1}^N {\mathbb Z} s_i$. At least in our case, $L(\Phi)$ includes $\Phi$. 
It implies that every root is a sum $\sum_{i=1}^N a_i s_i$. In these cases, the coefficients $a_i$ are positive integers for each positive root.  
For two vectors ${\boldsymbol x}=(x_1, \cdots , x_N), {\boldsymbol y}=(y_1 , \cdots , y_N)$ we obtain an inner product $({\boldsymbol x}, {\boldsymbol y})= x_1 y_1 + \cdots + x_N y_N$. 
Let us define a matrix ${\cal C}_{ij} = 2 (s_i, s_j)/(s_j,s_j)$. 
\begin{align}
{\cal C}[A_3] &= \begin{pmatrix}  
2 & -1 & 0 \\ -1 & 2 & -1 \\ 0 & -1 & 2
\end{pmatrix} ~,&
{\cal C}[B_3] &= \begin{pmatrix}  
2 & -1 & 0 \\ -2 & 2 & -1 \\ 0 & -1 & 2 
\end{pmatrix} ~,&
{\cal C}[C_3] &= \begin{pmatrix}  
2 & -2 & 0 \\ -1 & 2 & -1 \\ 0 & -1 & 2 
\end{pmatrix} ~,&
{\cal C}[D_3] &= \begin{pmatrix}  
2 & 0 & -1 \\ 0 & 2 & -1 \\ -1 & -1 & 2 
\end{pmatrix} ~.
\end{align}
These matrices are equal to Cartan matrices(We will omit the definition of Cartan matrix).  

Their determinants are $\det{\cal C}[A_N]=N+1, \det{\cal C}[B_N] =\det{\cal C}[C_N]=2, \det{\cal C}[D_N]=4$ (You can check these values with physical induction). They are orders of  centers of corresponding simply connected groups. 

\begin{itemize}
\item $A_N$ :~ $s_i$ $(1\leq i \leq N)$, $s_j + s_{j-1} + \cdots + s_i$ $(1 \leq i < j \leq N)$~;
\item $B_N$ :~ $s_i$ $(1\leq i \leq N)$, $s_j + s_{j-1} + \cdots + s_i$ $(1 \leq i < j \leq N)$~,\\
~~~~~ $s_j+ \cdots  +s_{i+1} + 2 s_{i} + \cdots + 2 s_1$ $(1 \leq i < j \leq N)$~;
\item $C_N$ :~ $s_i$ $(1\leq i \leq N)$, $s_j + s_{j-1} + \cdots + s_i$ $(1 \leq i < j \leq N)$~,\\
~~~~~ $2 s_{i} + \cdots + 2s_2 + s_1$ $(2 \leq i \leq N)$,\\
~~~~~ $s_j + \cdots + s_{i+1}+ 2 s_{i} + \cdots + 2s_2 + s_1$ $(2 \leq i < j \leq N)$~;
\item $D_N$ :~ $s_i$ $(1 \leq i \leq N)$, $s_j + s_{j-1} + \cdots + s_i$ $(1 \leq i < j \leq N-1)$~,\\
~~~~~ $s_j + s_{j-1} + \cdots + s_3 + s_1$ $(3 \leq j \leq N)$,$s_j + s_{j-1} + \cdots + s_3 +s_2 + s_1$ $(3 \leq j \leq N)$,\\
~~~~~ $s_j + s_{j-1} + \cdots + s_{i+1} + 2 s_i + \cdots + 2s_3 + s_2 + s_1$ $(3\leq i <  j \leq N)$
\end{itemize}

 Weyl groups for $A,B,C,D$ series are  
\begin{align}
W[A_{N-1}] &= {\mathfrak S}_N~,&
W[B_{N}] &= {\mathfrak S}_N \rtimes \left( {\mathbb Z}/2 {\mathbb Z} \right)^N ~,&
W[C_{N}] &= {\mathfrak S}_N \rtimes \left( {\mathbb Z}/2 {\mathbb Z} \right)^N~,&
W[D_{N}] &= {\mathfrak S}_N \rtimes \left( {\mathbb Z}/2 {\mathbb Z} \right)^{N-1}~,
\end{align}

\section{Van der Monde determinant}
In this section, we consider the measure on the Lie algebra consisting of Hermitian matrices. 
We start from the basics of Lie algebra $sl(n)$. 
\begin{align}
{\mathfrak g} &:= \{ X \in M(n,\mathbb C ) |  {\rm Tr}(X)=0  \}~,& {\mathfrak h} &:= \{ H \in {\mathfrak g} | H ={\rm diag} (h_1, h_2 , \cdots , h_n) \}~. 
\end{align}
Elements belonging to ${\mathfrak h}$ commute with each other. We define adjoint action of $X \in {\mathfrak g}$ on $Y\in {\mathfrak g}$: ${\rm ad} (X) (Y):= [X,Y]$. For $X , Y \in {\mathfrak g}$, $[{\rm ad}(X) , {\rm ad}(Y)] = {\rm ad}([X,Y]) $. This is the adjoint representation of ${\mathfrak g}$. 
Especially, for $H_1 , H_2 \in {\mathfrak h}$, $[{\rm ad}(H_1) , {\rm ad}(H_2)] = 0$ and there are simultaneous eigenvectors $T\in {\mathfrak g}$ of ${\mathfrak h}$, where $T$ is not an element in ${\mathfrak h}$. Let $T$ be such a generator, {\it i.e.} for arbitrary $H \in {\mathfrak h}$, there exists a complex number $\lambda$ such that $[H , T] = \lambda T$. Suppose that for $H_1, H_2 \in {\mathfrak h}$, the eigenvalues are $\lambda_1 , \lambda_2 \in \mathbb C$: $[H_1 , T] = \lambda_1 T , [H_2 , T ] = \lambda_2 T$. 
\begin{align}
[a_1 H_1 + a_2 H_2 , T] &= ( a_1 \lambda_1 + a_2 \lambda_2)  T~, 
\end{align}
This means that there is a linear map $\psi$ from ${\mathfrak h}$ to the complex number field $\mathbb C$: $[ H , T ] = \psi(H) T$, $H \in {\mathfrak h}$. 
The dual space of ${\mathfrak h}$ is defined as ${\mathfrak h}^{\vee} := {\rm Hom}({\mathfrak h} , \mathbb C)$. The linear map $\psi$ belongs to ${\mathfrak h}^{\vee}$. 
We call $\psi$ root. 
Let us define the set consisting of eigenvectors: ${\mathfrak s}:=\{ T \in {\mathfrak g} \setminus {\mathfrak h} | \exists \alpha \in {\mathfrak h}^{\vee} , \forall H \in {\mathfrak h}, [H,T] = \alpha(H) T   \}$. We define a set consisting of all roots: 
$\Phi := \{ \alpha \in {\mathfrak h}^{\vee} | [ H , T] =  \alpha(H) T , T \in {\mathfrak s} \}$. Because ${\mathfrak g}$ has finite dimensions, $\Phi$ should be a finite set. $\Phi$ is called root system. Let $\alpha$ be a root. The eigenspace with respect to $\alpha$ is
\begin{align}
V_{\alpha} &:= \{ X \in {\mathfrak g} | \forall H \in {\mathfrak h}, ~ [H,X] = \alpha(H) X \}~.
\end{align} 
The dimension of this space is one. 
We define an inner product on ${\mathfrak g}$,  
$\kappa(X,Y) := {\rm Tr} ( {\rm ad}(X) {\rm ad}(Y) )$. This inner product induces an inner product $( \cdot , \cdot )$ on ${\mathfrak h}$. 
\begin{align}
( H , H ) &:= {\rm Tr} ({\rm ad} (H) {\rm ad}(H)) = \sum_{\alpha \in \Phi} \alpha(H)^2
\end{align}
Because ${\mathfrak g}$ is semi-simple Lie algebra, this inner product $( H,H)$ is nondegenerate. 
We can construct orthonormal basis of this space:  $\{ {\boldsymbol e}_i \}_{i=1,\cdots,n}$, $( {\boldsymbol e}_i , {\boldsymbol e}_j ) = \delta_{ij}$. This inner product induces an isomorphism between ${\mathfrak h}$ and ${\mathfrak h}^{\vee}$. 
\begin{align}
\varphi &: {\mathfrak h} \rightarrow {\mathfrak h}^{\vee}~,& \varphi(H_1)(H_2):= (H_1,H_2)~.  
\end{align}

Suppose that $X$ is a Hermitian matrix. A measure on this space is defined as follows
\begin{align}
\prod_{i=1}^{n} dX_{ii} \prod_{i \leq j} d\Re{X_{ij}}d\Im{X_{ij}}  \propto  \prod_{i=1}^n d h_i \prod_{\alpha \in \Phi} d X_{\alpha}
\end{align}
A unitary matrix $U$ diagonalize the matrix $X$ like $X= U\Lambda U^{-1}$, where $\Lambda$ is a diagonal matrix. 
Let us consider its differentials
\begin{align}
d X &= d( U\Lambda U^{-1} ) = U(d \Lambda ) U^{-1} + (d U) \Lambda U^{-1} + U \Lambda ( d U^{-1} )~,\cr
&= U \left( d \Lambda - {\rm ad}(\Lambda) (   {\cal A} )    \right) U^{-1}
\end{align}
where the differential form ${\cal A} = U^{-1}(dU)$ takes value in the Lie algebra and the components along to the Cartan algebra do not affect. Therefore this form is defined on $U(n)/T^n$, where $T^n$ is a maximal torus of $U(n)$. The  adjoint action of ${\rm ad}(\Lambda)$ yields the van-der Monde determinant term as the Jacobian: $\det|_{{\rm root}} ( {\rm ad}(\Lambda) ) = \prod_{\alpha : \mbox{root}} (\alpha , \Lambda )$. 
We can easily generalize this result to other groups.

\section{Evaluation of integrals and special functions}
Here we present our evaluation of the residue calculus. 
\begin{align}
A_{N-1}&: \phi_i = (N+1)/2-i,& 
B_{N}&: \phi_i= N-i+1,&
C_N&: \phi_i = N-i+1/2,&
D_N&: \phi_i = N-i,
\end{align}
where $1 \leq i \leq N$. 

$\phi_i - \phi_j = (j-i)$. 
\begin{align}
{\cal Z}[A_{N-1}] &= \frac{1}{N!} \left( \prod_{j=3}^N \prod_{i=1}^{j-2} \frac{j-i}{j-i-1} \right) \left( \prod_{j=2}^N \prod_{i=1}^{j-1} \frac{j-i}{j-i+1} \right)\\
&= \frac{1}{N!}  \left( \prod_{j=3}^N \frac{2 \cdot 3 \cdots (j-1)}{1 \cdot 2 \cdots (j-2)} \right) \left( \prod_{j=2}^N 
\frac{1 \cdot 2 \cdots (j-1)}{2 \cdot 3 \cdots j} \right)\\
&= \frac{1}{N!}  \left( \prod_{j=3}^N (j-1) \right) \left( \prod_{j=2}^N 
{j} \right)^{-1}= \frac{1}{N \cdot N!}~.
\end{align}
$\phi_i - \phi_j = (j-i)$, $\phi_i + \phi_j = 2N-(i+j)$.
\begin{align}
{\cal Z}[D_{N}] &= \frac{1}{2^{N-1} N \cdot  N!} \left( \prod_{j=2}^{N-1} \prod_{i=1}^{j-1} \frac{2N-(i+j)}{(2N-1)-(i+j)} \right)
\left( \prod_{i=1}^{N-2} \frac{N-i}{N-1-i} \right)
\left( \prod_{j=2}^N \prod_{i=1}^{j-1} \frac{2N-(i+j)}{(2N+1)-(i+j)} \right)\\
&= \frac{N-1}{2^{N-1} N \cdot  N!}  \left( \prod_{j=2}^{N-1} \frac{2N-j-1}{2N-2j} \right)
\left( \prod_{j=2}^N \frac{2N-2j+1}{2N-j} \right)\\
&= \frac{1}{2^N N \cdot  N!} \frac{(2N-3)!! }{(2N-4)!!}~.
\end{align}
$\phi_i - \phi_j = (j-i)$, $\phi_i+\phi_j = 2(N+1) -(i+j)$, $\phi_i = N-i+1$. 
\begin{align}
{\cal Z}[B_{N}] &= \frac{1}{2^{N} N \cdot  N!} \left( \prod_{j=2}^{N} \prod_{i=1}^{j-1} \frac{(2N+2)-(i+j)}{(2N+1)-(i+j)} \right)
\left( \prod_{j=2}^N \prod_{i=1}^{j-1} \frac{(2N+2)-(i+j)}{(2N+3)-(i+j)} \right)\\
&\times \left( \prod_{i=1}^{N-1} \frac{(N+1)-i}{N-i} \right)\left( \prod_{i=1}^{N} \frac{(N+1)-i}{(N+2)-i} \right)\\
&= \frac{1}{2^{N} N \cdot  N!} \left( \prod_{j=2}^{N} \frac{2N-j+1}{2N-2j+2} \right)
\left( \prod_{j=2}^N  \frac{2N-2j+3}{2N-j+2} \right)\frac{N}{N+1}\\
&= \frac{1}{2^{N} N \cdot  N!} 
\frac{(2N-1)!!}{(2N-2)!!}
\frac{N+1}{2N}\frac{N}{N+1}\\
&= \frac{1}{2^{N+1} N \cdot  N!} 
\frac{(2N-1)!!}{(2N-2)!!}
\end{align}

$\phi_i - \phi_j = (j-i)$, $\phi_i+\phi_j = (2N+1) -(i+j)$, $2\phi_i = (2N+1)-2i$. 
\begin{align}
{\cal Z}[C_{N}] &= \frac{1}{2^{N} N \cdot  N!} \left( \prod_{j=2}^{N} \prod_{i=1}^{j-1} \frac{(2N+1)-(i+j)}{2N-(i+j)} \right)
\left( \prod_{j=2}^N \prod_{i=1}^{j-1} \frac{(2N+1)-(i+j)}{(2N+2)-(i+j)} \right)\\
&\times \left( \prod_{i=1}^{N-1} \frac{(2N+1)-2i}{2N-2i} \right)\left( \prod_{i=1}^{N} \frac{(2N+1)-2i}{(2N+2)-2i} \right)\\
&=\frac{1}{2^{N} N \cdot  N!} \left( \prod_{j=2}^{N} \prod_{i=1}^{j-1} \frac{(2N+1)-(i+j)}{2N-(i+j)} \right)
\left( \prod_{j=2}^N \prod_{i=1}^{j-1} \frac{(2N+1)-(i+j)}{(2N+2)-(i+j)} \right)\\
&\times \left( \prod_{i=1}^{N-1} \frac{(2N+1)-2i}{2N-2i} \right)\left( \prod_{i=1}^{N} \frac{(2N+1)-2i}{(2N+2)-2i} \right)\\
&=\frac{1}{2^{N} N \cdot  N!} \left( \prod_{j=2}^{N} \frac{2N-j}{2N-2j+1} \right)
\left( \prod_{j=2}^N \frac{2N-2j+2}{2N-j+1} \right)
 \frac{(2N-1)!!}{(2N-2)!!}  \frac{(2N-1)!!}{(2N)!!} \\
 &= \frac{1}{2^{N} N \cdot  N!} \frac{(2N-2)!!}{(2N-3)!!} \frac{N}{2N-1} \frac{(2N-1)!!}{(2N-2)!!}  \frac{(2N-1)!!}{(2N)!!}\\
  &= \frac{1}{2^{N+1} N \cdot  N!}    \frac{(2N-1)!!}{(2N-2)!!}
\end{align}
These numbers lead us to several special functions.  
The series
\begin{align}
\Phi(s , t ; z ) &= 1 + \frac{s}{t} \frac{z}{1!} + \frac{s (s +1)}{t (t +1)} \frac{z^2}{2!} + 
\frac{s (s +1)(s +2)}{t (t +1)(t +2) } \frac{z^3}{3!} + \cdots
\end{align}
is called a {\it confluent hypergeometric function}. The product $s (s+1) \cdots (s+n-1)$ is denoted $(s)_n$. 
The function $\Phi(s,t;z)$ is a solution of the differential equation 
\begin{align}
z\frac{d^2 F}{dz^2} + (\gamma -z) \frac{dF}{dz} - \alpha F &=0~.  
\end{align}

\begin{align}
{\rm Ei}(x) &:= - \lim_{\epsilon \rightarrow +0} \left[  \int_{-x}^{-\epsilon} \frac{e^{-t}}{t}dt + \int_{\epsilon}^{\infty} \frac{e^{-t}}{t} dt  \right] = {\rm PV} \int_{- \infty}^x \frac{e^t}{t} dt~,~~(x>0)\\
&= {\bf C} + \ln x + \sum_{k=1}^{\infty} \frac{x^k}{k \cdot k!}~,
\end{align}
where ${\bf C}$ is the Euler constant; 
\begin{align}
{\bf C} = \lim_{n \rightarrow \infty} \left( \sum_{k=1}^{n-1} \frac{1}{k} - \ln n \right)  = 0.577~215~664~90 \cdots
\end{align}
The asymptotic behavior of the function ${\rm Ei}(x)$ with respect to the limit $x \rightarrow + \infty$ is
\begin{align}
{\rm Ei}(x) &= \frac{e^x}{x} \left[ \sum_{k=0}^n \frac{k!}{z^k} + R_n(x) \right]~,&
|R_n(x)| = O(|z|^{-n-1})~
\end{align}

The series 
\begin{align}
{}_p F_q ( s_1,s_2, \cdots, s_p; t_1 , t_2 , \cdots , t_q; z) &= \sum_{k=0}^{\infty} \frac{(s_1)_k (s_2)_k \cdots (s_p)_k}{(t_1)_k (t_2)_k \cdots (t_q)_k} \frac{z^k}{k!}
\end{align}
is called a {\it generalized hypergeometric series}. 
The confluent hypergeometric function is one of the generalized hypergeometric series;
$\Psi(s,t;z) = {}_1 F_1 (s;t;z)$. There are several recursion relations and ladder operators. 
\begin{align}
\frac{z}{\gamma} \Phi(\alpha +1, \gamma +1;z) &= \Phi(\alpha +1, \gamma ;z) - \Phi(\alpha , \gamma ;z)~,&
\frac{d}{dz} \Phi(\alpha , \gamma ;z) &= \frac{\alpha}{\gamma} \Phi(\alpha +1, \gamma +1 ;z)~,
\end{align}
\begin{align}
\left( z \frac{d}{dz} + \alpha  \right) \Phi( \alpha  , \gamma ; z) &= \alpha \Phi(\alpha +1 , \gamma ;z ) 
\end{align}

\end{document}